 \documentclass{emulateapj}

\usepackage[backref,breaklinks,colorlinks,citecolor=blue]{hyperref}
\usepackage[all]{hypcap} 

\usepackage{color}

\newcommand{\cluster}{SDSS~J1110+6459}

\newcommand{\giantarc}{SGAS~J111020.0+645950.8}
\newcommand{\hst}{\textit{HST}}

\newcommand{\zlens}{0.659}
\newcommand{\zA}{2.481}

\newcommand{\msol}{\mathrm{M_\odot}}

\shorttitle{\cluster}
\slugcomment{ApJ in prep: draft date \today}
\shortauthors{Johnson et al.}

\begin{document}

\title{STAR FORMATION AT $z=\zA$ IN THE LENSED GALAXY SDSS J1110+6459, I: \\Lens Modeling and Source Reconstruction}

\author{
Traci L. Johnson\altaffilmark{1}, Keren Sharon\altaffilmark{1}, Michael D. Gladders\altaffilmark{2,3}, Jane R. Rigby\altaffilmark{4}, Matthew B. Bayliss\altaffilmark{5}, Eva Wuyts\altaffilmark{6}, Katherine E. Whitaker\altaffilmark{7,8}$^\dagger$, Michael Florian\altaffilmark{2}, Katherine T. Murray\altaffilmark{9}
}
\email{tljohn@umich.edu}
\altaffiltext{*}{Based on observations made with the NASA/ESA Hubble Space Telescope, obtained at the Space Telescope Science Institute, which is operated by the Association of Universities for Research in Astronomy, Inc., under NASA contract NAS 5-26555. These observations are associated with program \# 13003.}
\altaffiltext{1}{University of Michigan, Department of Astronomy, 1085 South University Avenue, Ann Arbor, MI 48109, USA}
\altaffiltext{2}{Department of Astronomy \& Astrophysics, The University of Chicago, 5640 S. Ellis Avenue, Chicago, IL 60637, USA}
\altaffiltext{3}{Kavli Institute for Cosmological Physics at the University of Chicago, USA}
\altaffiltext{4}{Observational Cosmology Lab, NASA Goddard Space Flight Center, Greenbelt MD 20771, USA}
\altaffiltext{5}{Kavli Institute for Astrophysics \& Space Research, Massachusetts Institute of Technology, 77 Massachusetts Ave., 
Cambridge, MA 02139, USA}
\altaffiltext{6}{ArmenTeKort, Antwerp, Belgium}
\altaffiltext{7}{Department of Astronomy, University of Massachusetts--Amherst, Amherst, MA 01003, USA}
\altaffiltext{8}{Department of Physics, University of Connecticut, Storrs, CT 06269, USA}
\altaffiltext{9}{Space Telescope Science Institute, 3700 San Martin Drive, Baltimore, MD 21218, USA}
\altaffiltext{$\dagger$}{Hubble Fellow}
\begin{abstract}
Using the combined resolving power of the {\it Hubble Space Telescope} and gravitational lensing, we resolve star-forming structures in a $z\sim2.5$ galaxy on scales much smaller than the usual kiloparsec diffraction limit of \hst. \giantarc\ is a clumpy, star forming galaxy lensed by the galaxy cluster \cluster\ at $z=\zlens$, with a total magnification $\sim30\times$ across the entire arc. We use a hybrid parametric/non-parametric strong lensing mass model to compute the deflection and magnification of this giant arc, reconstruct the light distribution of the lensed galaxy in the source plane, and resolve the star formation into two dozen clumps. We develop a forward-modeling technique to model each clump in the source plane. We ray trace the model to the image plane, convolve with the instrumental point spread function (PSF), and compare with the GALFIT model of the clumps in the image plane, which decomposes clump structure from more extended emission. This technique has the advantage, over ray tracing, by accounting for the asymmetric lensing shear of the galaxy in the image plane and the instrument PSF. At this resolution, we can begin to study star formation on a clump-by-clump basis, toward the goal of understanding feedback mechanisms and the buildup of exponential disks at high redshift.
\end{abstract}


\section{Introduction}

Through surveys of galaxies over cosmic time, we now know that the peak era of star formation in galaxies occurred around $z=2$, with half of the stars observed today being formed by $z=1.3$ \citep[][and references therein]{Madau:2014qd}. Cold dark matter overdensities collapse to form halos onto which cold gas can accrete in the form of filaments \citep{Keres:2005yq,Genzel:2006rt}, fueling star formation and thus making these halos highly efficient stellar factories \citep{Behroozi:2013vn}. It is thought that gravitational instabilities within the gaseous disk collapse to form stars \citep{Toomre:1964fr,Dekel:2006qf,Brooks:2009jk}, which give these galaxies clumpy surface brightness distributions, the predecessors to the exponential disk galaxies of today \citep{Elmegreen:2005xy,Elmegreen:2007qv,Elmegreen:2009nr,Forster-Schreiber:2011rz,Forster-Schreiber:2011sf,Guo:2011rm,Guo:2015zl}. They are the launching points for feedback-driven outflows \citep{Genzel:2008gf,Genzel:2011ys}, which can be powerful enough to eject metals from the galaxy, potentially provide the gas needed to harbor future star formation, and may migrate inward and coalesce to form the bulges of spiral galaxies. Understanding the properties of these star forming clumps provides insight into the growth and content of galaxies at $z=0$.

The {\it Hubble Space Telescope} (\hst) can resolve galactic structure on the kiloparsec scale at intermediate redshifts (the resolution limit of \hst\ is $\sim530$ pc at $z=1$ at rest-frame optical wavelengths). The typical size (projected full width half maximum; FWHM) of clumps in high-redshift galaxies found in \hst\ imaging are reported to be $\sim1$ kpc \citep{Elmegreen:2007qv,Forster-Schreiber:2011rz,Livermore:2012fk}. Even the largest stellar complexes in the local universe hardly reach these sizes \citep{Kennicutt:1984mz}; we expect these clumps at high redshift to be mostly unresolved with the best telescopes available today. However, gravitational lensing can overcome these resolution limits, as the magnification increases the overall area of the source, allowing us to probe scales less than 100 parsecs in extremely bright, highly magnified galaxies \citep{Jones:2010uq,Swinbank:2010zr,Livermore:2012fk,Livermore:2015ve}.

Here, we measure the physical sizes of star forming regions of the galaxy \giantarc, a giant arc at $z=\zA$ lensed by the galaxy cluster \cluster\ at $z=\zlens$. This arc is one of the most striking in a larger sample of strongly lensed giant arcs, described in \S~\ref{sec:sgas}, and has also been found in other strong lensing cluster searches \citep{Stark:2013kl}. As we will show in \S~\ref{sec:lensmodel}, the arc is composed of three merging images with a total magnification of $28\pm8$. \hst\ Wide Field Camera 3 (WFC3) imaging in UVIS and IR has revealed here that this lensed galaxy is speckled with clumpy structure near the scale of the \hst\ PSF.

We discuss \hst\ observations and follow-up spectroscopy of the lensing system in \S~\ref{sec:observations}. In \S~\ref{sec:lensmodel}, we discuss our method of strong lens modeling, which involves a new hybrid parametric/non-parametric technique developed specifically for this cluster, as it shows complex mass structure requiring more flexibility than traditional parametric lens modeling methods. We have also developed a forward modeling technique for modeling the clumpy structure within the giant arc in the source plane, and then ray tracing this model to the image plane, as we describe in detail in \S~\ref{sec:clump_model}. The source plane model provides us with a picture of the galaxy delensed and deconvolved from the PSF, from which we can, for the first time, measure physical properties on physical scales well less than 100 parsecs at this redshift. We summarize our measurements of clump luminosity and size of this galaxy in the source plane in \S~\ref{sec:summary}. Finally, in \S~\ref{sec:conclusion}, we summarize our work and discuss our plans for extending the methods of this study to a larger sample of high-redshift, high-magnification lensed galaxies in the future.

Throughout this paper, we assume a flat cosmology with $\Omega_M=0.3$, $\Omega_\Lambda=0.7$, and $H_0=70\ \mathrm{km\ s^{-1}\ Mpc^{-1}}$, for which an angular size of 1\arcsec\ corresponds to a physical distance of 6.97 kpc at the cluster redshift $z=\zlens$ and 8.085 kpc at the redshift of the giant arc $z=\zA$. All magnitudes are reported in the AB system.

\section{The Sloan Giant Arc Survey and \giantarc}
\label{sec:sgas}

\giantarc\ was discovered as part of the Sloan Giant Arcs Survey (SGAS; Gladders et al., in preparation), a program systematically searching for strong lensing galaxy clusters in the Sloan Digital Sky Survey \citep[SDSS; ][]{York:2000sp}. Galaxy clusters were selected from the SDSS Data Release 7 photometric catalog using a red sequence cluster-finding algorithm \citep[e.g.,][]{Gladders:2000kq}. SDSS images in $g$, $r$, $i$, and $z$ were combined into custom color images spanning $4'\times4'$ around each cluster center, with scale parameters selected to allow the best contrast for visually detecting faint extended features. The images were visually inspected and ranked by our team, and instances of strong lensing features were noted. Lower-confidence lens candidates were targeted for imaging follow-up by larger telescopes (e.g., Gemini and Magellan). The highest-confidence lensed galaxies and those confirmed through follow-up imaging were targeted for spectroscopy \citep{Bayliss:2011ul,Bayliss:2011gf,Bayliss:2012rt}, confirming hundreds of lenses, with a well-understood completeness and purity of the survey.  

\section{Observations}
\label{sec:observations}

\subsection{Hubble Space Telescope}
As part of the extensive SGAS follow-up campaign, we obtained \hst\ imaging of 37 SGAS clusters, which strongly lens over 70 background sources (\hst\ Cycle~23, GO13003, PI Gladders). As part of this program, \cluster\ was imaged with the \hst/WFC3 on 2013 January 8 UT over three orbits, using four broadband filters: F105W (1112 s) and F160W (1212 s) in the infrared (IR) channel, and F390W  (1212 s) and F606W (2420 s) in the UVIS channel. The selected filters span the broadest possible wavelength space accessed by HST with good sensitivity, with particular filters chosen to provide clean sampling of the age-sensitive D4000 break. 

The imaging within each filter consists of four sub-pixel dither positions required for point spread function (PSF) reconstruction, cosmic ray rejection, and chip gap compensation. The IR data were taken using the SPARS25 readout sequence mode. Each exposure was reduced with the WFC3 data-reduction pipeline, combined with the Astrodrizzle routine \citep{Fruchter:2010lr}, and for each filter, drizzled onto a common grid with a pixel scale of 0\farcs03 and drop sizes of 0\farcs08 and 0\farcs05 for UVIS and IR, respectively. We experimented with different pixel scales and drop sizes, and found that this combination provides the best sampling of the PSF. The IR channel contains circular areas of decreased sensitivity, referred to as the ``IR blobs" in the WFC3 Data Handbook \citep{Deustua:2016ab}. We developed a custom algorithm for removing these artifacts by modeling each ``IR blob" with GALFIT \citep{Peng:2010qy} for each observation in our SGAS program and then combining all models into a superflat frame. Each observation was flat-fielded with this frame prior to drizzling.

The UVIS channel suffers declining charge transfer efficiency (CTE), which can cause large flux decreases and higher correlated readout noise. To mitigate these losses, our UVIS F390W observations were taken with post-flash to increase the background level and ensure that the lowest surface brightness sources had high enough counts \citep{Rajan:2010ab}. We used the Pixel-based Empirical CTE Correction Software\footnote{\url{http://www.stsci.edu/hst/wfc3/ins\_performance/CTE/}} provided by STScI to apply post-observation image corrections to the individual exposures. The reduced data set yields a 5$\sigma$ limiting magnitude of $m= 26.43, 26.47, 25.36, \mathrm{and}\ 25.68$ mag with a 0\farcs7 diameter aperture in F390W, F606W, F105W, and F160W, respectively.

\subsection{Spitzer/IRAC}

Data from the IRAC instrument of the {\it Spitzer} Space Telescope, obtained during the post-cryogenic ``warm mission," were as follows.  Shallow 3.6 and 4.5 ~\micron\ images were obtained in Cycle 7 (program 70154, PI M.~Gladders); much deeper 3.6~\micron\ images were obtained in Cycle 9 (program 90232, PI J.~Rigby). We combine data from both programs. The average per-pixel integration time, excluding field edges, was 11.7 ks at 3.6~\micron, and 1.14 ks at 4.5~\micron.

We reduced the \textit{Spitzer} IRAC data by following the general guidance of the IRAC Cookbook\footnote{\url{http://irsa.ipac.caltech.edu/data/SPITZER/docs/irac/iracinstrumenthandbook/}} for reducing the COSMOS medium-deep data, albeit with more stringent (3~$\sigma$) outlier rejection, as well as residual bias correction. We started with the corrected basic calibrated data products (cBCDs) from the \textit{Spitzer} archive. We applied the warm mission column pulldown correction (\texttt{bandcor\_warm} by Matt Ashby). Because residual bias pattern noise and persistence can dominate over the background in deep integrations, we constructed images of the residual bias, also known as a ``delta dark frame.''  For each channel in each observation, a residual bias correction was created from all the cBCDs, by detecting and masking sources in each image, adjusting the pedestal offset level of each image so that the modes had the same value, and then taking the median with 3~$\sigma$ outlier rejection.  The relevant median image was then subtracted from every cBCD image in that channel and that observation. For each filter, individual images were combined into a mosaic using the Mopex command-line tools. We used the overlap correct tool to add an additive correction for each residual-bias-corrected cBCD image to bring it to a common sky background level.  These images were then combined into a mosaic using the Mopex mosaic tool, using the drizzle algorithm with a pixel fraction of 0.6, and 3~$\sigma$ outlier rejection using the box outlier rejection method.

\subsection{Gemini/Gemini Mulit-Object Spectrograph (GMOS)}
\label{sec:specz}
Spectroscopic observations for the field of \cluster\ were taken with the GMOS \citep{Hook:2004yq} on the {\it Gemini North} telescope as part of queue programs GN-2011A-Q-19 (PI:~Gladders) and GN-2015B-Q-26 (PI:~Sharon). Two custom multi-object nod and shuffle slit masks were designed, one for each program, targeting both lensed galaxies and candidate cluster members using the R400 grism with the OG515 order blocking filter, following the design described in \citet{Bayliss:2011ul}. The first (second) slit mask was observed for $2\times40$ min on 2012 March 29 (2015 January 8) with seeing 0\farcs66 (1\farcs09) and airmass 1.42-1.45 (1.47-1.42).

\begin{figure}
\includegraphics[scale=0.4,resolution=300]{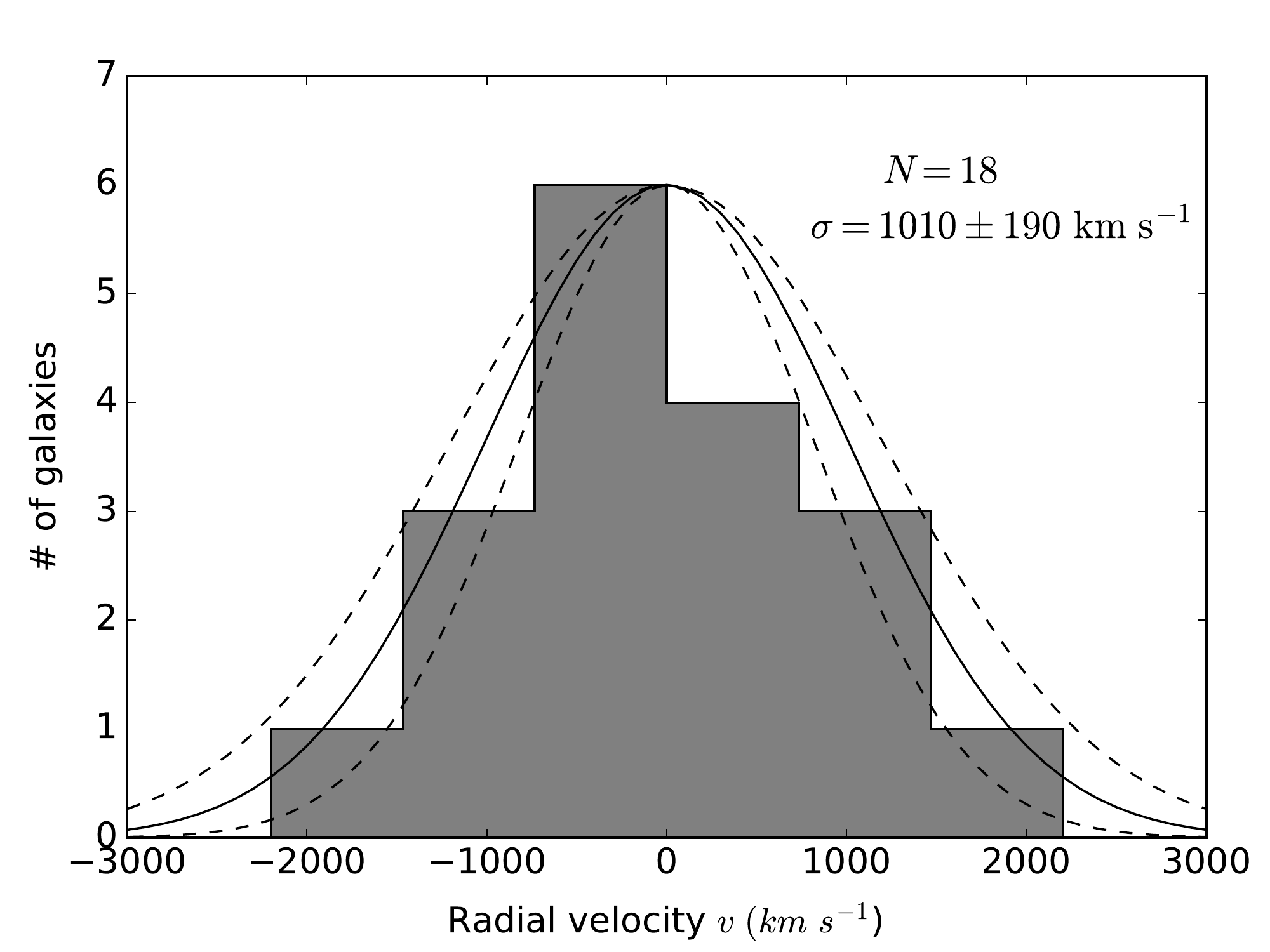}
\caption{Histogram of the radial velocities of the spectroscopically confirmed cluster member galaxies in \cluster\ with respect to the bi-weight center at $z=0.656$. We overplot Gaussians centered on the bi-weight center with widths set to the velocity dispersion (solid line) and its $1\sigma$ errors (dashed lines).}
\label{fig:vel_hist}
\end{figure}

We list all the spectroscopic redshifts from the GMOS observations in \autoref{tab:redshifts}. We spectroscopically confirm 17 of these galaxies as cluster members with $0.64<z<0.67$. The redshift of the brightest cluster galaxy (BCG), $z=0.659$, was measured independently by \citet{Oguri:2012bs} and \citet{Stark:2013kl}. With $N=18$ galaxies, we can obtain a rough estimate of the dynamical mass of \cluster\ from the radial velocity dispersion~$\sigma$. We use the bi-weight average and spread from \citep{Beers:1990rt} to estimate the central (average) redshift of the cluster and its velocity dispersion. We determine a cluster central redshift $z=0.656$ and velocity dispersion $\sigma=1010\pm190\ \mathrm{km\ s^{-1}}$. We use a jackknife to estimate the radial velocity measurements. \autoref{fig:vel_hist} shows a histogram of the radial velocities of galaxies in \cluster.


\capstartfalse
\begin{deluxetable}{cccc}
\tablecolumns{5}
\tablecaption{\cluster\ spectroscopically confirmed cluster members and other objects}
\tablehead{\colhead{R.A.} & \colhead{Decl.} & \colhead{$z$} & \colhead{Distance from} \\
	\colhead{(J2000)} & \colhead{(J2000)} &  & \colhead{BCG (\arcsec)}}
\startdata
11:10:08.81 & +65:00:35.5 & $0.6559\pm0.0005$ & 73.89\\
11:10:11.53 & +65:00:03.7 & $0.6547\pm0.0004$ & 42.36\\
11:10:11.90 & +64:58:21.7 & $0.6631\pm0.0007$ & 93.78\\
11:10:13.01 & +65:00:17.6 & $0.6587\pm0.0008$ & 42.15\\
11:10:13.03 & +64:59:28.6 & $0.6495\pm0.0004$ & 35.47\\
11:10:16.36 & +64:59:22.2 & $0.6610\pm0.0010$ & 27.10\\
11:10:17.23 & +64:59:27.9 & $0.6550\pm0.0004$ & 20.21\\
11:10:17.56 & +64:59:38.9 & $0.6501\pm0.0003$ & 9.02\\
11:10:17.73 & +64:59:47.9 & $0.659$\tablenotemark{*} & 0.00\\
11:10:18.45 & +64:59:37.5 & $0.6677\pm0.0010$ & 11.37\\
11:10:18.48 & +64:59:52.7 & $0.6606\pm0.0002$ & 6.79\\
11:10:18.50 & +64:59:58.8 & $0.6447\pm0.0004$ & 11.92\\
11:10:18.51 & +65:00:40.7 & $0.6523\pm0.0007$ & 53.01\\
11:10:18.92 & +64:59:47.7 & $0.6557\pm0.0010$ & 7.59\\
11:10:21.16 & +65:00:39.0 & $0.6562\pm0.0003$ & 55.54\\
11:10:22.06 & +64:58:29.9 & $0.6490\pm0.0004$ & 82.73\\
11:10:23.37 & +64:59:24.4 & $0.6585\pm0.0003$ & 42.84\\
11:10:24.70 & +65:00:24.4 & $0.6560\pm0.0008$ & 57.28\\
\hline
11:10:08.60 & +64:59:32.4 & $1.2480\pm0.0010$ & 59.90\\
11:10:12.03 & +64:58:35.7 & $0.3392\pm0.0001$ & 80.79\\
11:10:12.49 & +64:58:41.2 & $0.7551\pm0.0005$ & 74.53\\
11:10:14.88 & +64:58:36.9 & $0.7518\pm0.0006$ & 73.25\\
11:10:19.55 & +64:59:58.3 & $2.4801\pm0.0010$ & 15.51\\
11:10:19.97 & +64:59:44.4 & $2.4817\pm0.0010$ & 14.64\\
11:10:19.99 & +64:59:44.4 & $2.4808\pm0.0020$ & 14.79\\
11:10:19.99 & +64:59:51.0 & $2.4807\pm0.0025$ & 14.70\\
11:10:30.71 & +65:00:40.9 & $0.5495\pm0.0002$ & 97.90
\enddata
\tablenotetext{*}{From \citet{Oguri:2012bs} and \citet{Stark:2013kl}.}
\label{tab:redshifts}
\end{deluxetable}
\capstarttrue

\begin{figure*}
\includegraphics[scale=0.48,resolution=300]{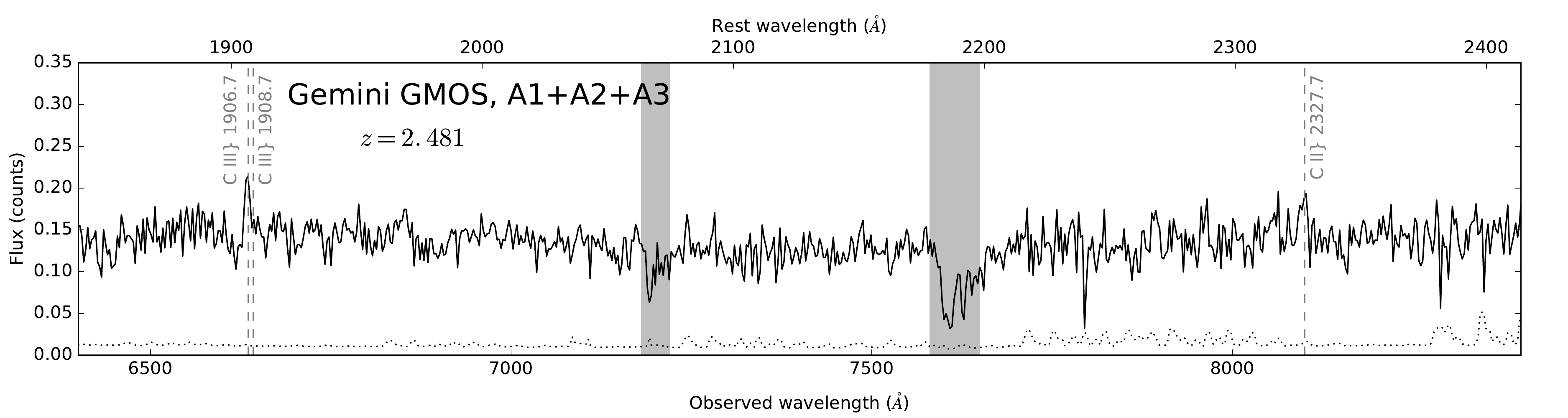}
\caption{\textit{Gemini} GMOS spectrum of \giantarc, summed from slits placed on all three images (A1, A2, and A3). The dotted line indicates the noise level, and the gray bands indicates part of the spectrum with strong telluric absorption. The vertical gray-dashed lines indicate the locations of rest-frame UV emission lines.}
\label{fig:GMOSspectrum}
\end{figure*}

\autoref{fig:GMOSspectrum} shows the summed spectrum of \giantarc; we determine a redshift of ($z = 2.4812 \pm0.0005$) from the summed spectrum of four slits placed on the giant arc covering all three images, derived from C~II] and C~III] nebular emission lines. Also visible in the spectra are several low-ionization ISM absorption lines with a systemic redshift of $2.480\pm 0.001$, corresponding to a $\sim100\ \mathrm{km\ s^{-1}}$ outflow.

We targeted two of these lensed galaxies, which were identified as strong lensing constraints (see \S~\ref{subsec:sec_arcs}), in the \textit{Gemini} Fast Turnaround (FT) program GN-2015A-FT-15 (PI: Johnson, 4.75 hr) using GMOS in long-slit observing mode. Observations were made with the B600 grism and the 1\farcs5-width long slit, with the slit positioned to target images B1, B2, and several other objects. The final spectra include a total integration time of 9000 s, resulting in spectra covering a wavelength range, $\Delta \lambda \sim 4150-6970$\AA. The spectra of both lensed galaxies include low S/N continuum flux, but no strong features that enable a redshift measurement.

In both the GMOS MOS observations (2011) and FT long-slit, we detect emission from a star-forming galaxy located near B1 (shown in \autoref{fig:arclabels}). From both observations, we confirm a redshift of $z=0.6447$, based on [OII] 3727\AA\ and Balmer lines for this galaxy, confirming it as a cluster member. Based on its characteristic morphology, this galaxy can be classified as a jellyfish galaxy -- cluster member galaxies with jellyfish-like morphology that exhibit trails of knotted star formation as they pass through the hot intercluster medium and are stripped of their cold gas \citep{Ebeling:2014rf,Suyu:2010xy}.

\subsection{MMT/Blue Channel Spectrograph}
\begin{figure*}
\includegraphics[scale=0.48,resolution=300]{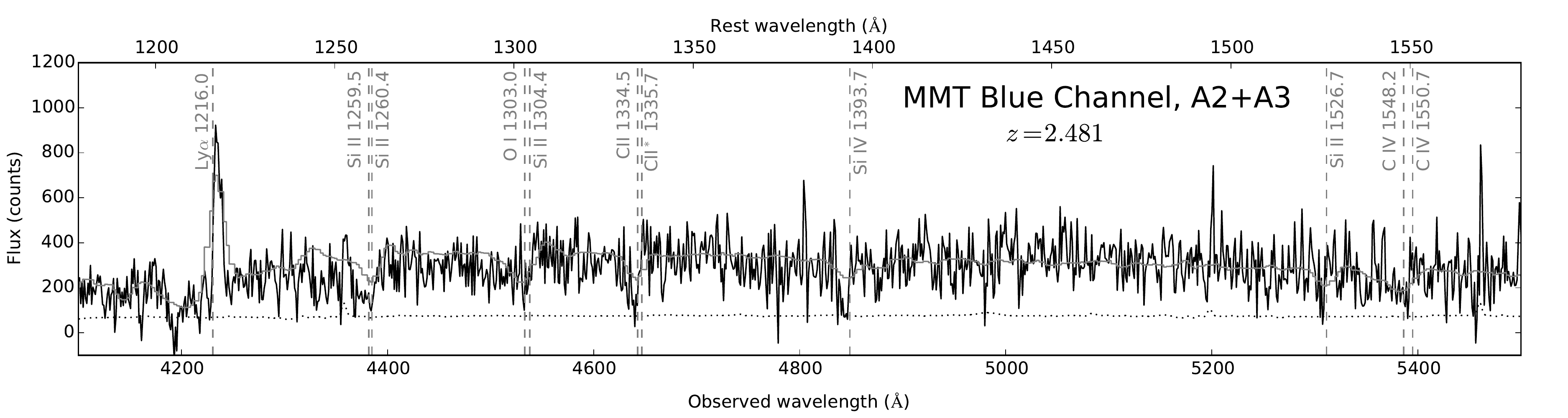}
\caption{MMT Blue Channel Spectrograph spectrum of \giantarc. The dotted line indicates the noise level, and the gray-solid line is a spectral template of Ly$\alpha$-emitting galaxies with strong absorption features at $z\sim3$ galaxies from \citet{Shapley:2003fk}. The vertical gray-dashed lines indicate the locations of rest-frame UV emission and absorption lines.}
\label{fig:MMTspectrum}
\end{figure*}

\giantarc\ was observed on 2015 May 5 with the Blue Channel spectrograph on the 6.5m MMT telescope at Mt. Hopkins, AZ. The spectrograph was configured with a 1\farcs25\ wide longslit and the 500 line mm$^{-1}$ grating, resulting in a dispersion of 1.19~\AA\ per pixel, and a spectral resolution, $\delta \lambda \simeq 4.1$~\AA. The data cover a total range in wavelength, $\Delta\lambda = 4000-7150$~\AA. We acquired a total integration time of 6000 s (two 3000 s exposures), and the longslit was aligned along the length of the arc, resulting in emission that extends $\sim$15\arcsec\ along the slit. We measure $z = 2.481$ from numerous features that are common in the rest-UV spectra of starburst galaxies, including Ly$\alpha$ emission and absorption from low-ionization species of Si, C, and O (as shown in \autoref{fig:MMTspectrum}). We note that a spectroscopic redshift for \giantarc\ was reported by \citet{Stark:2013kl} and agrees with our value.

\section{Strong Lensing Analysis of \cluster}
\label{sec:lensmodel}

\subsection{Previous lensing analysis}
\citet{Oguri:2012bs} use ground-based imaging from the {\it Subaru} telescope to compute strong lensing and weak lensing mass models of \cluster. The strong lens model is severely under-constrained; the primary arc structure could not be resolved, and the source redshift of the primary arc had not yet been spectroscopically confirmed (assumed $z=2\pm1$). Although the secondary arcs we use in our model (which we will discuss in \S~\ref{subsec:sec_arcs}) are clearly visible in the Subaru imaging, they were not identified or used as constraints in the model. The \citet{Oguri:2012bs} model, with a single mass component, can adequately estimate the mass within the Einstein radius for the fiducial redshift assumed for the redshift of the primary arc. \citet{Oguri:2012bs} note that the weak lensing map suggests the presence of a more complicated mass distribution than indicated from strong lens modeling.

\subsection{Lensing evidence}
With the improved resolution of \hst, we include additional structure within the giant arcs and faint secondary image systems as additional constraints, allowing for a more complex lens model of this cluster. Additionally, the spectroscopic redshifts we have obtained for this cluster help break the mass-sheet degeneracy \citep{Schneider:1995vn} and constrain the slope of the mass distribution.

\begin{figure*}
\centering
\includegraphics[height=0.32\textheight]{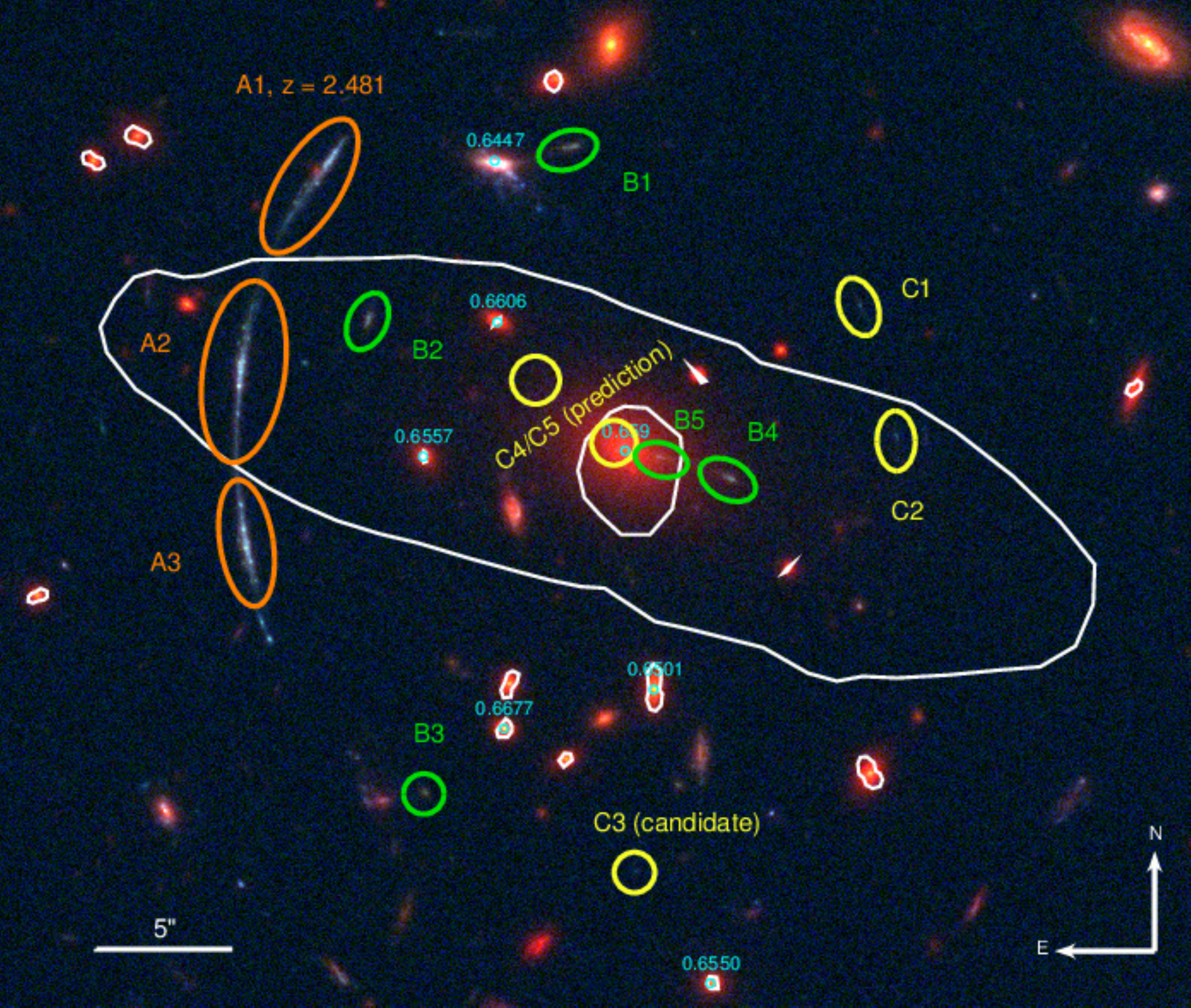}
\includegraphics[height=0.32\textheight]{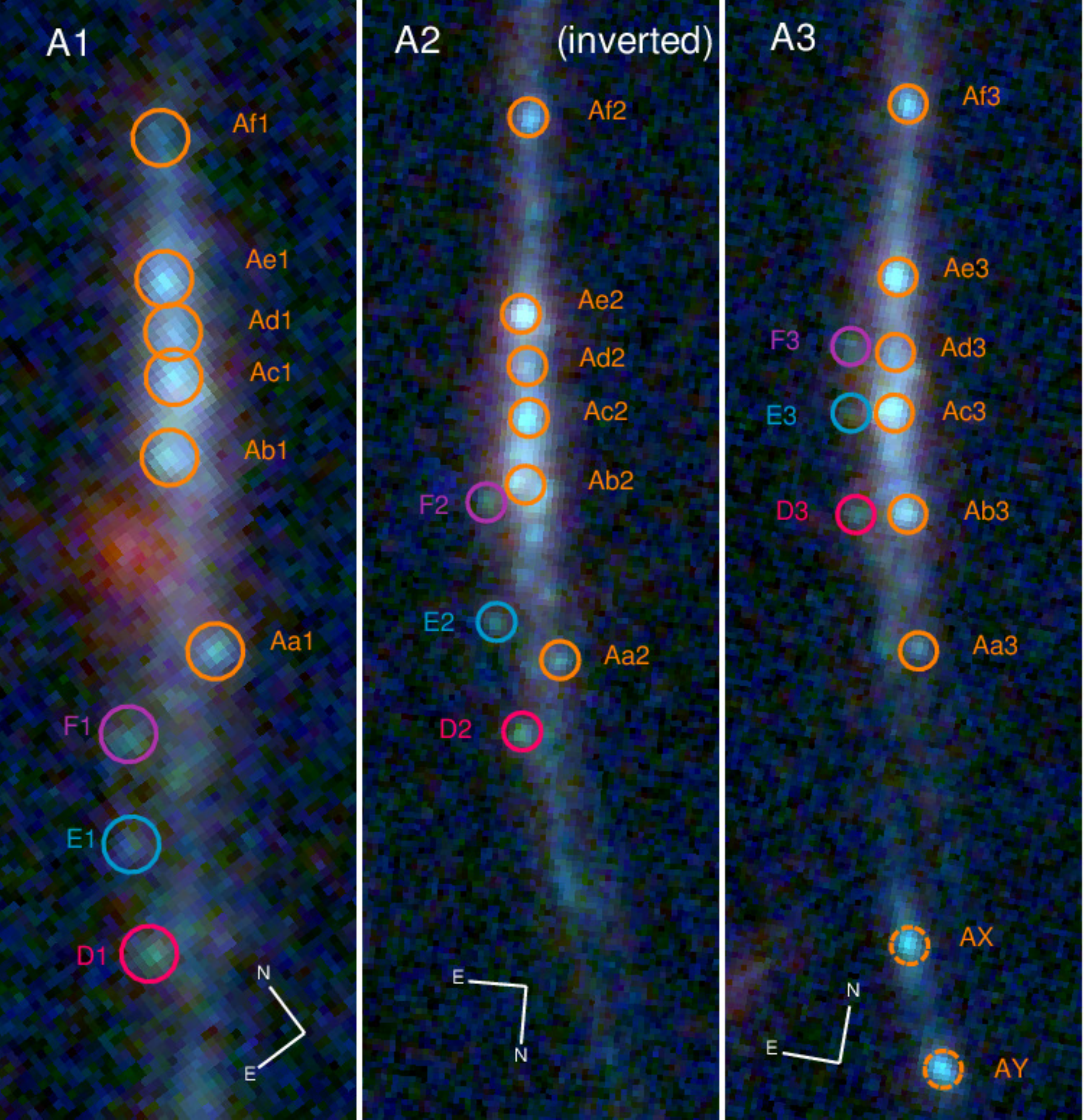} \\ \vspace{10pt}
\includegraphics[width=0.91\textwidth]{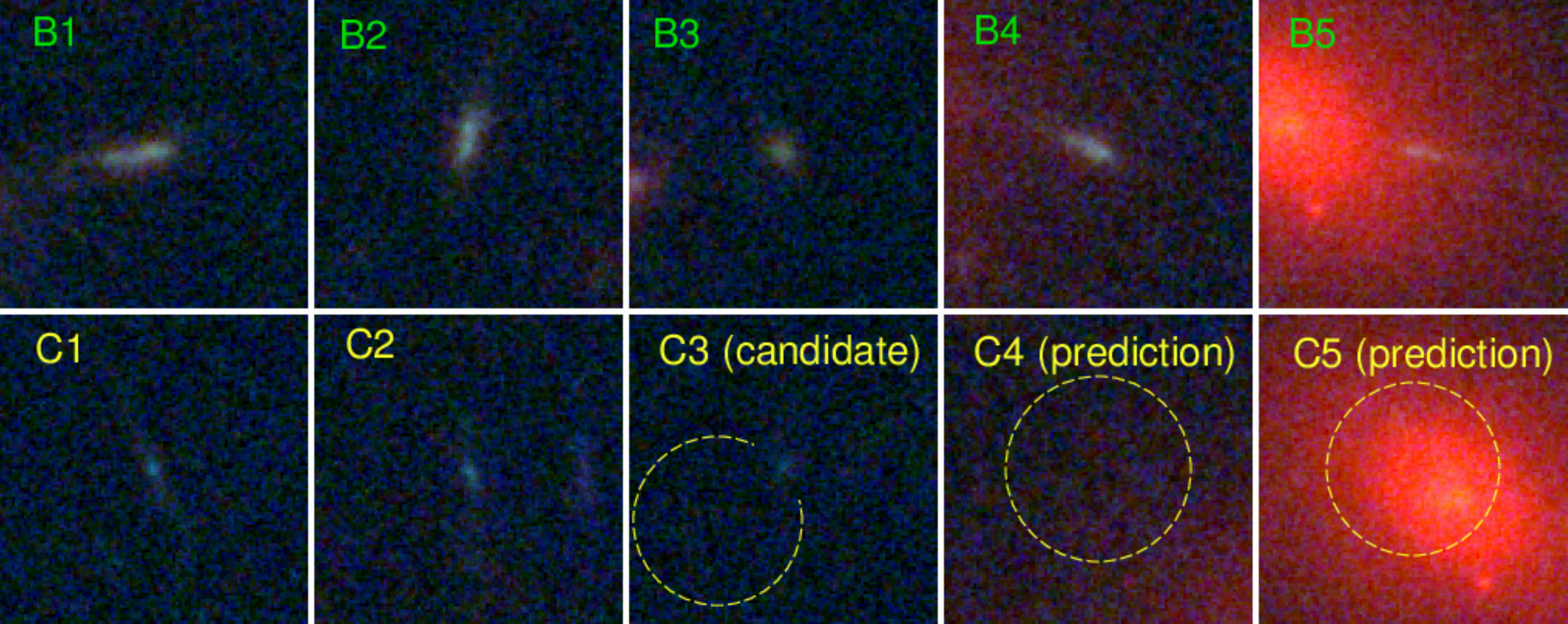}
\caption{(Top left) \hst\ WFC3 imaging of \cluster\ in F160W (red), F606W (green), and F390W (blue). Labeled are image systems A, B, and C used in the lens modeling. The redshifts of other objects from \autoref{tab:redshifts} are shown in cyan. The critical curve for $z=\zA$ is shown by the white lines. (Top right) Close up image of the three images of the main arc A and systems D, E, and F; the middle image has been inverted along the N-S direction to match parity with the other two images. The clumps labeled A[a-f] are individual clumps matched across all three images used as constraints in the model. AX and AY are likely part of the lensed galaxy, but lie outside the caustic region and thus are not multiply imaged. Each circle is 0\farcs1 in radius. (Bottom) Images of systems B and C. The dashed circle indicates the center and rms scatter of the predicted images marginalized over all models which predict that image. Each postage stamp cutout is 3\arcsec x3\arcsec.}
\label{fig:arclabels}
\end{figure*}

We identify three unique sources, lensed into a total of 11 images by \cluster, and use the positions of 10 of these images as constraints on the lens model (see \autoref{tab:arcs}). The constraint positions are centered on distinct morphological or chromatic features of the galaxy that are seen in each image. This means that the method is best done by eye rather than a quantitative identifier (i.e. peak emission or barycenter), especially given that the magnifications of these features can vary dramatically from image to image. We include a positional error of 0\farcs3 in the image positions to account for possible small-scale deflections due to structure or galaxy lensing. The positions of the clump features of the giant arc are included as additional constraints. We show the positions of the image constraints in \autoref{fig:arclabels} and list their coordinates in \autoref{tab:arcs}. 

\capstartfalse
\begin{deluxetable}{cccc}
\tablecaption{Identifications of lensed arcs in \cluster}
\tablehead{\colhead{Arc ID} & \colhead{R.A.}       & \colhead{Dec.}      & \colhead{Model$^{b}$} \\
	         \colhead{} &            \colhead{(J2000)} & \colhead{(J2000)} &  \colhead{$z$}}
\startdata
Aa1 & 11:10:19.56 & +64:59:57.88 & 2.481$^{a}$ \\
Aa2 & 11:10:19.96 & +64:59:52.06 &  $\dots$ \\
Aa3 & 11:10:19.92 & +64:59:42.98 &  $\dots$ \\
Ab1 & 11:10:19.51 & +64:59:58.53 & $\dots$ \\
Ab2 & 11:10:20.00 & +64:59:51.16 & $\dots$ \\
Ab3 & 11:10:19.94 & +64:59:43.69 & $\dots$ \\
Ac1 & 11:10:19.48 & +64:59:58.75 & $\dots$ \\
Ac2 & 11:10:20.00 & +64:59:50.81 & $\dots$ \\
Ac3 & 11:10:19.97 & +64:59:44.21 & $\dots$ \\
Ad1 & 11:10:19.47 & +64:59:58.88 & $\dots$ \\
Ad2 & 11:10:20.01 & +64:59:50.54 & $\dots$ \\
Ad3 & 11:10:19.98 & +64:59:44.53 & $\dots$ \\
Ae1 & 11:10:19.45 & +64:59:59.05 & $\dots$ \\
Ae2 & 11:10:20.02 & +64:59:50.27 & $\dots$ \\
Ae3 & 11:10:19.99 & +64:59:44.93 & $\dots$ \\
Af1 & 11:10:19.41 & +64:59:59.46 & $\dots$ \\
Af2 & 11:10:20.03 & +64:59:49.24 & $\dots$ \\
Af3 & 11:10:20.00 & +64:59:45.85 & $\dots$ \\
AX$^c$ & 11:10:19.78 & +64:59:40.85 & $\dots$ \\
AY$^c$ & 11:10:19.83 & +64:59:41.51 & $\dots$ \\
\hline
B1 & 11:10:19.25 & +64:59:52.61 & $3.79\pm0.17$ \\
B2 & 11:10:18.03 & +64:59:59.27 & $\dots$ \\
B3 & 11:10:18.91 & +64:59:35.06 & $\dots$ \\
B4 & 11:10:17.10 & +64:59:46.80 & $\dots$ \\
B5 & 11:10:17.54 & +64:59:47.63 & $\dots$ \\
\hline
C1 & 11:10:16.34 & +64:59:53.33 & $3.82\pm0.24$ \\
C2 & 11:10:16.14 & +64:59:48.46 & $\dots$ \\
C3$^d$ & 11:10:17.674 & +64:59:32.10 & $\dots$ \\
C3$^e$ & 11:10:17.775 & +64:59:31.59 & $\dots$ \\
C4$^e$ & 11:10:18.257 & +64:59:50.54 & $\dots$ \\
C5$^e$ & 11:10:17.796 & +64:59:48.20 & $\dots$ \\
\hline
D1 & 11:10:19.69 & +64:59:57.16 & $2.39\pm0.02$ \\
D2 & 11:10:19.99 & +64:59:52.45 & $\dots$ \\
D3 & 11:10:19.99 & +64:59:43.64 & $\dots$ \\
\hline
E1 & 11:10:19.66 & +64:59:57.51 & $2.37\pm0.02$ \\
E2 & 11:10:20.02 & +64:59:51.89 & $\dots$ \\
E3 & 11:10:20.01 & +64:59:44.17 & $\dots$ \\
\hline
F1 & 11:10:19.62 & +64:59:57.83 & $2.35\pm0.03$ \\
F2 & 11:10:20.03 & +64:59:51.27 & $\dots$ \\
F3 & 11:10:20.02 & +64:59:44.52 &  $\dots$
\enddata
\tablenotetext{$^a$}{Redshift of system A is fixed to the spectroscopic redshift.}
\tablenotetext{$^b$}{The model redshifts are marginalized over all eight lens models.}
\tablenotetext{$^c$}{AX and AY are part of A3, but are not multiply imaged and are not used as constraints in the model.}
\tablenotetext{$^d$}{This galaxy was identified as a possible counter image of system C but was not used as a constraint in the lens model.}
\tablenotetext{$^e$}{Predicted image locations, marginalized over all the models for which an image was predicted.}
\label{tab:arcs}
\end{deluxetable}
\capstarttrue

\subsubsection{Primary arc \giantarc}

The primary arc \giantarc\ stretches $\sim$17\arcsec\ in length, and is $\sim$15\arcsec\ from the BCG. It consists of three images that partially merge together in the image plane (labeled A in \autoref{fig:arclabels}) with several bright emission knots visible in the \hst\ imaging. The center image, A2, has the highest magnification and most resolved structure. Many of the clumps identified within A2 are unresolved in the other images with lower magnification. With this in mind, we identify six groups of clumps that are multiply imaged, rather than the individual clumps, and use these groupings as model constraints.

There are two bright blobs slightly south of A3 that are likely part of the primary arc (labeled AX and AY in \autoref{fig:arclabels}). This portion of the galaxy containing these blobs lies outside of the caustic region, and therefore is not multiply imaged.

\subsubsection{Identification of multiply imaged galaxies}
\label{subsec:sec_arcs}
The identification of secondary lensed galaxies is done iteratively, by eye, with the help of the lens model. We identify five sets of multiply imaged secondary arcs (B--F in \autoref{fig:arclabels} and \autoref{tab:arcs}) in the \hst\ data based on image configuration, morphology, and color. The redshifts of these two background galaxies have not been spectroscopically confirmed, despite our best efforts with \textit{Gemini}/GMOS (see \S~\ref{sec:specz}), so we leave the redshifts of these secondary arcs as free parameters to be optimized in the lens model. We use estimates of the photometric redshifts (see \S~\ref{subsec:photoz}) for these galaxies as priors.

Arcs B1 and B2 are tangential arcs located 11\farcs6 north and 10\farcs9 northeast of the BCG, respectively. A third image B3 was predicted and discovered 14\farcs8 southeast of the BCG. We also identified the radial arcs B4 and B5 from color and morphology extending west 4\farcs1 and 1\farcs7 from the center of the BCG.

The faint pair of tangential arcs C1 and C2 are located 10\farcs2 northwest and 10\farcs1 west of the BCG, respectively. We also find a possible candidate for a third image C3 15\farcs8 south of the BCG that matches in color. The location of this candidate is consistent with the image configuration; however, there is a large statistical error on the predicted location of the third image, due to the uncertainty in the redshift of this image system. Also, this image is predicted further from the critical curve than C1 and C2, and has a lower magnification and tangential shear observable by eye, making it difficult to confirm this candidate as the third image by shape. Therefore, we do not include this candidate image as a constraint in the lens modeling.

Near to the primary arc are three bright specks that are slightly different in color than the main arc and are also triply imaged in the same configuration. During the lens modeling process, we found that the positions of these blobs are not as well-reconstructed as the clumps used as constraints within the giant arc when fixed at the same source redshift, suggesting that this may be a separate system at a different redshift. We therefore include these three image systems (D, E, and F) in the lens model, with their redshifts as free parameters.

For all images without spectroscopic redshifts, we use a uniform random prior of $1<z<5$ for the free parameter in the lens model.

The model predicted redshifts for all secondary arcs B--F are listed in \autoref{tab:arcs}. For image system B, the model predicts a much higher redshift than the photometric prediction.

\subsubsection{Photometric Redshifts of Secondary Arcs}
\label{subsec:photoz}

\begin{figure*}
\includegraphics[scale=0.8,resolution=300,trim=0 150 0 150]{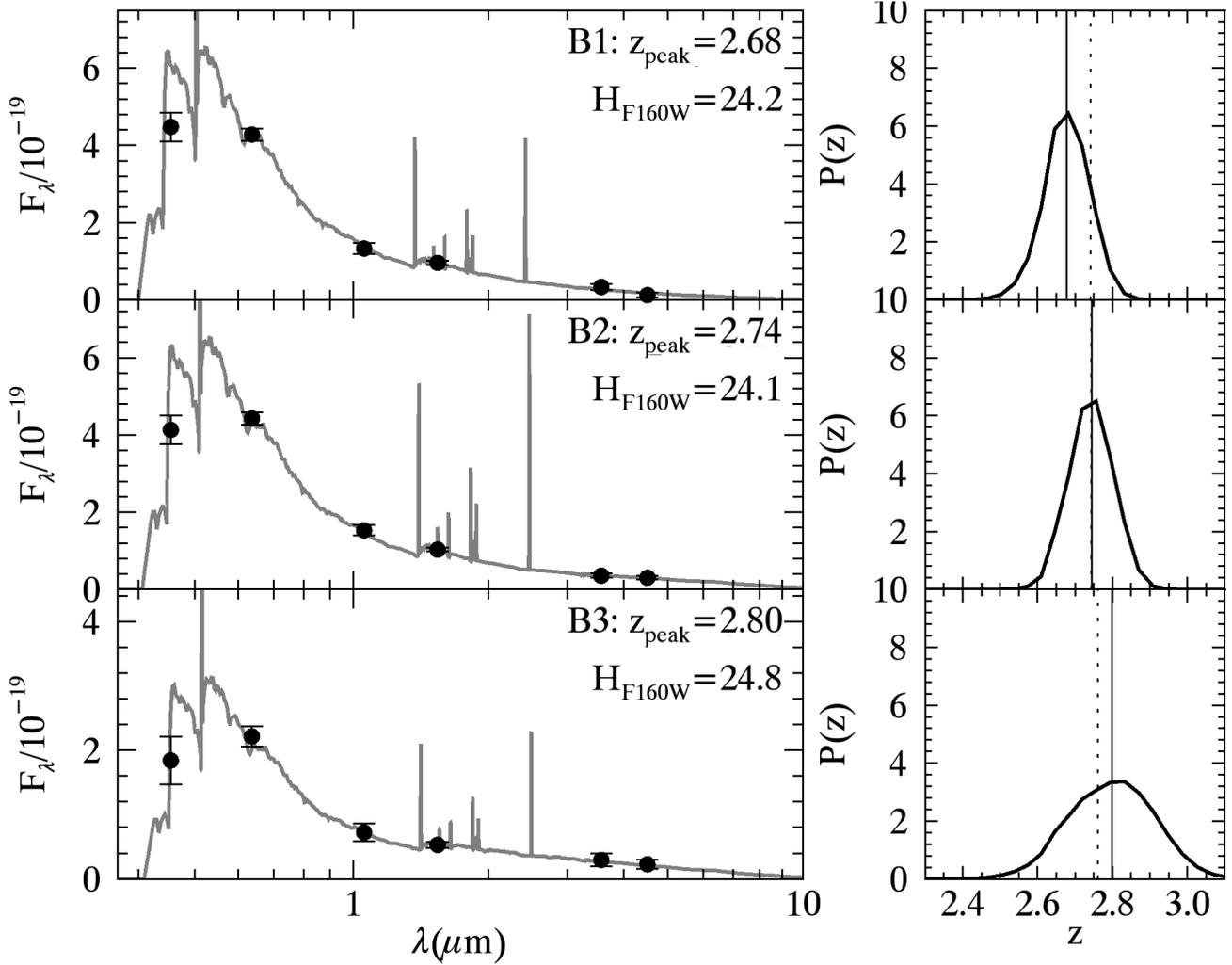}
\caption{Spectral energy distribution and photometric redshift for arcs B1, B2, and B3. $z_\mathrm{peak}$ is the highest-probability $z$ weighted by $P(z)$: $z_\mathrm{peak} = \int zp(z)dz/\int P(z) dz$. $F_\lambda$ is in units of erg s$^{-1}$ cm$^{-2}$ \AA$^{-1}$.}
\label{fig:photoz}
\end{figure*}

The photometry of all objects in the \hst\ and {\it Spitzer} imaging was extracted following procedures outlined in \citet{Skelton:2014lr}. Spectral energy distribution (SED) fits and photometric redshifts were computed for all objects using all four \hst\ filters, and the two \textit{Spitzer} IRAC channels using the EAZY redshift code \citep{Brammer:2008uq}. \autoref{fig:photoz} shows the photometric redshifts of secondary arcs B1, B2, and B3. The redshift probability distribution functions (PDF) support the identification of these images as images of the same galaxy at $z\sim2.7$ ($z_{peak} = 2.68, 2.74, 2.80$, respectively).  At the photometric redshift of image B, this would imply Ly$\alpha$ at $\sim$4500\AA, which was not detected in the GMOS spectrum reported in \S~\ref{sec:specz}. The lack of such emission does not preclude this photometric redshift, however, as many star-forming galaxies have little or no Ly$\alpha$ emission. There are no other potentially strong emission lines located in the bandpass of the GMOS spectra for this photo-$z$, and so we conclude that the existing spectral data are consistent with the photometric analysis. C1 and C2 were also extracted for photometry for the \hst/UVIS filters but were undetected in Spitzer/IRAC -- four filters were not enough to extract a robust photometric redshift for this image system. Images D, E, and F are all blended with the giant arc, especially in the IR bands, and thus could not be extracted for photometry. Consequently, they do not have photometric redshifts.

\subsection{Lens modeling process}
To compute the lens model of \cluster, we used the publicly available software \texttt{LENSTOOL} \citep{Jullo:2007lr}, which utilizes a Markov-Chain Monte Carlo (MCMC) to optimize the parameters of the lensing potential from Bayesian evidence. All of the components of the potential are modeled as pseudo-isothermal elliptical mass distributions \citep[PIEMD;][]{Limousin:2005cr}, which are described by seven parameters: a position $x$ and $y$; an ellipticity $e=(a^2-b^2)/(a^2+b^2)$ where $a$ and $b$ are the semi-major and semi-minor axes, respectively; a position angle $\theta$; a fiducial velocity dispersion $\sigma$; a core radius $r_\mathrm{core}$; and a cut radius $r_\mathrm{cut}$.

The lens modeling is done iteratively. We begin with a set of constraints and our initial guess for the mass distribution within the cluster. Using a preliminary model, we search for new image candidates, which get added as constraints to the model and allow for more free parameters to be included in the next iteration. The early iterations are completed using a source plane optimization. Ideally, optimization should be done in the image plane, as this is where the model constraints lie; however, computation in the source plane is a much faster process, and provides a quick approximation for the lens model. The best-fit models we present here are the final iteration computed under image plane optimization.

\subsubsection{Lens plane mass components}
The total mass distribution of \cluster\ can be characterized by a smooth component encompassing the bulk of the cluster mass, which is perturbed by smaller halos occupied by galaxies. We use a red sequence selection criterion to select cluster member galaxies \citep[i.e.,][]{Gladders:2000kq}. We use the F606W-F105W colors for selecting the galaxies in \cluster\ that best sample the 4000 \AA\ break at the cluster redshift. The galaxies lying on the red sequence are assigned a unique halo with the parameters determined by the \texttt{SExtractor} \citep{Bertin:1996ly} outputs for location, ellipticity, and position angle from the F105W image. We adhere to a light-traces-mass methodology for modeling the perturbing halos, in which brighter galaxies occupy a deeper potential well. The parameters that determine the total mass of the halo, i.e., velocity dispersion $\sigma_0$, core radius $r_\mathrm{core}$, and cut radius $r_\mathrm{cut}$, are scaled by the magnitudes in the F105W band following the relations in \citet{Jullo:2007lr},

\begin{eqnarray}
\sigma_0 &=&  \sigma_0^\star \left(\frac{L}{L^\star}\right)^{1/4} \\
r_\mathrm{core} &=& r_\mathrm{core}^\star  \left(\frac{L}{L^\star}\right)^{1/2} \\
r_\mathrm{cut} &=& r_\mathrm{cut}^\star  \left(\frac{L}{L^\star}\right)^{1/2},
\end{eqnarray}

\noindent where $\sigma_0,r_\mathrm{core},r_\mathrm{cut}$ are the parameters for an $L^\star$ galaxy. These scaling relations translate to a constant mass-to-light ratio for all of the cluster member galaxies. We determine the apparent magnitude of an $L_\star$ galaxy at $z=\zlens$ to be $m_\star=19.9$ in F105W, and we set $\sigma_\star=120\ \mathrm{km\ s^{-1}}$, $r_\mathrm{cut}^\star=30$ kpc, and $r_\mathrm{core}^\star=0.15\ \mathrm{kpc}$. These parameters can also be optimized in the modeling; however, we find that they cannot be constrained easily, as the individual galaxies have a very small and local effect on the lensing potential. Therefore, we choose to fix these parameters and apply deviations from this strict scaling law on individual galaxies when necessary. In this case, we chose to allow the velocity dispersions of the BCG and another galaxy located at R.A. = 11:10:557, decl. = +64:59:58.31 to be free parameters in the model. The BCG affects the slope of the inner mass distribution, and thus the positions of the radial arcs B4/B5. The second galaxy lies almost directly along the line of sight to image A1, and likely will cause small scale--but significant--perturbations to the lensing potential for the clumps in this image. We also use a circular lensing potential for this galaxy, because the flux from A1 interfered with extracting reasonable shape parameters.

We place a massive, cluster-scale halo (also PIEMD) near the location of the BCG. All the parameters are free to optimize, with the exception of the cut radius, which lies far outside the strong lensing regime. It cannot be constrained with the lensing images, so we fix the value to 1500 kpc.

\begin{figure*}
\center
\includegraphics[scale=0.7,resolution=300]{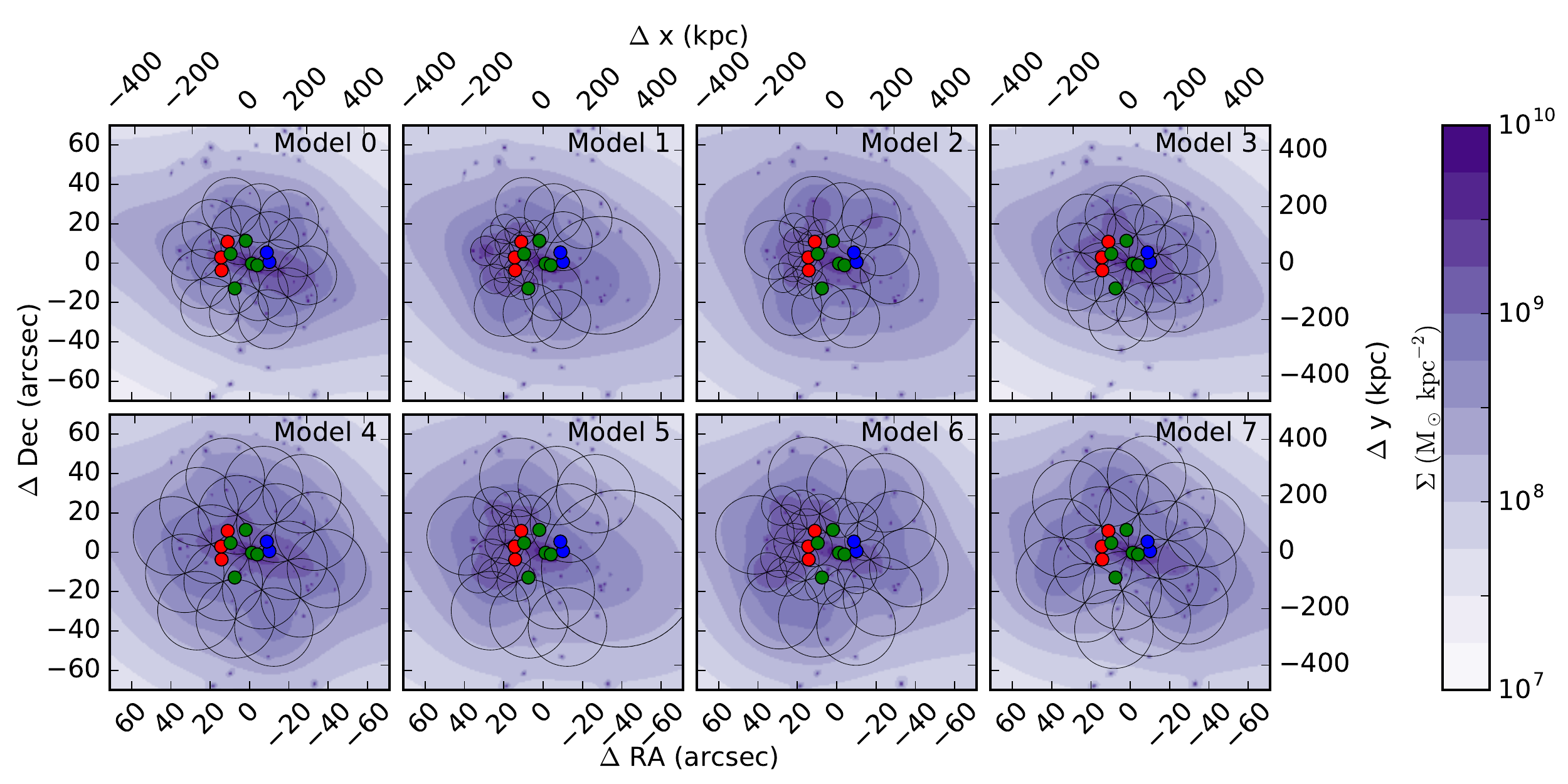}
\caption{Surface mass density of each lens model. The circles show where each node of the multi scale grid is centered, with radius equal to the cut radius of that node. Overlaid on grid for reference are the locations for arcs A (red), B (green), and C (blue). All offsets are given with respect to the center of the BCG.}
\label{fig:mass_distribution}
\end{figure*}

A parametric lens modeling approach is simplistic and appropriate when there are few lensing constraints in a model. However, cluster lensing systems are complicated by non-axisymmetric structure in the dark matter distribution and structures along the line of sight. In the case of \cluster, we find that the basic parametric model is insufficient for reconstructing the lensing, thus necessitating more flexibility. Specifically, models using only cluster-scale halos would, at best, produce an image plane rms of 1\farcs4. We develop a hybrid lens model for this cluster by adding a non-parametric multiscale grid component on top of the parametric cluster- and galaxy-scale halo components described above. We accomplish this via the following methods, developed by \citet{Jullo:2009ij}. We first construct a hexagonal-shaped grid within the lens plane with circular PIEMD halos, or nodes, located on the vertices and at the center, as shown in \autoref{fig:mass_distribution}. Each node forms an equilateral triangle with its adjacent nodes, and we set the cut radius equal to the side length of the triangle and set $r_\mathrm{cut}=(3/2)r_\mathrm{core}$. This parameterization is arbitrary, but was selected such that each grid halo is not cuspy--rather, each describes a perturbation in a largely smooth mass distribution. We only allow the velocity dispersion of each halo to vary. Thus, each node adds one additional free parameter to the entire lens model. Our over-constrained model allows for many more free parameters, so we allow for the inclusion of more nodes by recursively breaking up the grid into smaller fractal components. Each triangle of nodes split into four equilateral triangles of nodes, each of which is half the size of the original, and every node in the grid is set to the size of the smallest adjacent triangle. This process repeats for every triangle in the grid based on a specified node-breaking criteria and/or a maximum recursion depth. We exclude the nodes centered within a 12\arcsec radius from center of the BCG, where the massive cluster halo is located. We give the parameters for forming the multiscale grid in \autoref{tab:grid}. 

\capstartfalse
\begin{deluxetable*}{ccccccccc}
\tablecolumns{9}
\tablecaption{Multiscale grid parameters}
\tablehead{
	\colhead{} &
	\colhead{Grid Size} &
	\colhead{Grid P.A.} &
	\colhead{Grid Shift} &
	\colhead{Recursion} &
	\colhead{Threshold} &
	\colhead{\# of} &
	\colhead{Image Plane} &
	\colhead{Median Magnification} \\
	\colhead{} &
	\colhead{(\arcsec)} &
	\colhead{($^\circ$)} &
	\colhead{(\arcsec)} &
	\colhead{Depth} &
	\colhead{} &
	\colhead{Nodes} &
	\colhead{rms (\arcsec)} &
	\colhead{across arc}}
\startdata
Model 0 & 30 & -12 & 0 & 2 & 0 & 18 & 0.12 & 18 \\
Model 1 & 30 & -12 & 0 & 3 & 5 & 23 & 0.11 & 35 \\
Model 2 & 30 & -12 & 2.5 & 3 & 5 & 23 & 0.10 & 36 \\
Model 3 & 30 & -42 & 0 & 2 & 0 & 18 & 0.12 & 26 \\
Model 4 & 40 & -12 & 0 & 2 & 0 & 18 & 0.11 & 24 \\
Model 5 & 40 & -12 & 0 & 3 & 5 & 21 & 0.12 & 42 \\
Model 6 & 40 & -12 & 2.5 & 3 & 5 & 29 & 0.12 & 28 \\
Model 7 & 40 & -42 & 0 & 2 & 0 & 18 & 0.11 & 17
\enddata
\tablecomments{The grid size is the separation of the nodes and radius of the nodes at the first level of recursion depth. The position angle is the orientation of the major axis measured north of east ($-12^\circ$ aligns the major axis of the grid with the semi-major axis of the BCG). The grid shift is the shift of the center of the grid along the orientation of its major axis in the direction toward the middle image of arc A. The threshold is the number of constraints in each node required to recursively break that node into smaller nodes. The image plane rms is the rms scatter between the observed and predicted positions of the images used as constraints -- because all models use the same constraints, image plane rms serves as a measurement for goodness of fit.}
\label{tab:grid}
\end{deluxetable*}
\capstarttrue

Because our primary objective in adding more free parameters to the lens model is an accurate source magnification and reconstruction, we base our node-breaking criteria on the local density of constraints. For each triangle of nodes, we count the number of constraints within a circle connecting each of the three nodes, and the triangle is broken down into more nodes if the number density within the circle is above a set threshold. We present eight models using a variety of grid parameters -- node size, node-breaking criteria, recursion depth, grid center, and grid orientation. Each model produces similar masses and magnification. The image plane root mean square (rms) of all the models is $\sim0\farcs11$.

We estimate the uncertainties in the model parameters, magnifications, and masses from a suite of simulated models produced during the MCMC. We select models with the lowest image plane rms. Our model selection was chosen such that models span roughly the $1\sigma$ spread in values for the free parameters, which are well-constrained and have roughly Gaussian posterior probability distributions. This cut includes $\sim$100 models, and thus provides an adequate sampling of the parameter space around the best-fit model. Therefore, it can be used to estimate the statistical errors in the lens modeling.

\subsection{Mass distribution}

\begin{figure}
\includegraphics[scale=0.41,resolution=300]{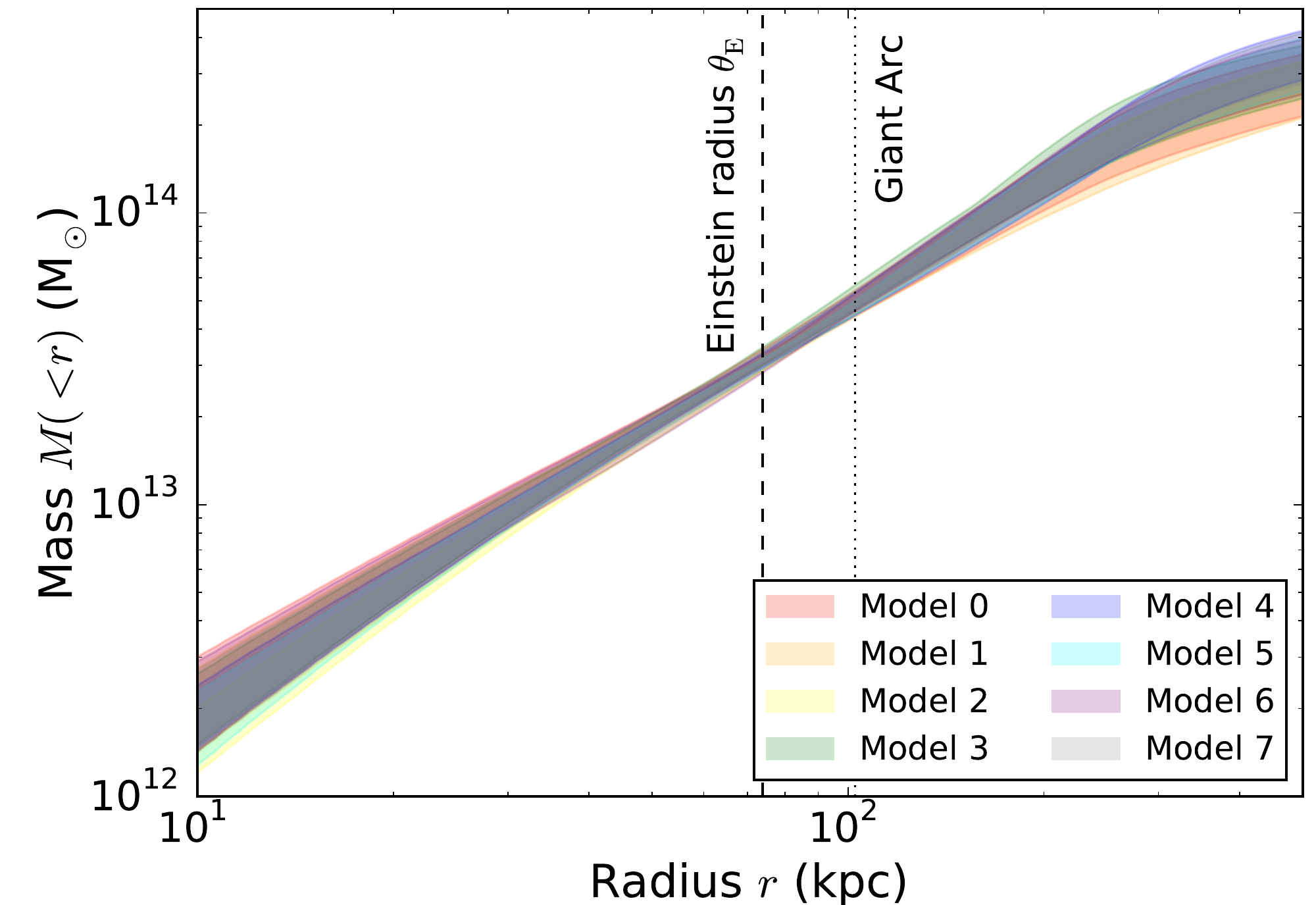}
\caption{We plot the integrated mass profile within radius $r$ of the galaxy cluster \cluster\ for each of the eight lens models. The shaded region indicates the $1\sigma$ uncertainty range in the mass. The dashed lines indicate the locations of the effective Einstein radius at $z=\zA$ and the radius of the giant arc. All models are in good agreement and converge at the Einstein radius, which corresponds roughly to the typical projected radius of the multiple images used as constraints in the lens models.}
\label{fig:mass_profile}
\end{figure}

The surface mass distribution for each of the eight lens model multiscale grid configurations is shown in \autoref{fig:mass_distribution}. The shape of the mass distribution beyond the location of the lensing constraints follows the design of the grid; however, the overall profile is well-constrained by the lensing. This is especially true on the eastern side of the cluster, in the vicinity of arc A, where the density of lensing constraints is high. We compute the integrated mass profile within radius $r$ of the galaxy cluster out to a radius of 500 kpc in \autoref{fig:mass_profile} for each of the eight models. The mass profiles of all models are the same within their statistical errors; they have especially good agreement around 100 kpc, approximately the projected radius of the strong lensing arcs used as constraints in the model (i.e., Einstein radius). Despite how much the surface mass density can change in regions when different parameterizations for the mass distribution are used, the total mass remains fairly robust.

Marginalizing over all eight models, we compute aperture masses centered on the BCG $M(r<250\ \mathrm{kpc}) = 1.7\pm0.1\times10^{14}\ \msol$ and $M(r<500\ \mathrm{kpc}) = 3.1\pm0.2\times10^{14}\ \msol$. The area enclosed within the $z=\zA$ critical curve is $A(<\mathrm{crit}) = 0.0983\pm0.003\ \square\arcmin$, enclosing a mass of $M(<\mathrm{crit}) = 3.3\pm0.1\times10^{13}\ \msol$. The effective Einstein radius for the giant arc is $\theta_\mathrm{E} = \sqrt{A(<\mathrm{crit})/\pi} = 10.6\pm0.2\arcsec$.

Although the models produce low statistical errors on the mass, we warn that the slope of the mass distribution is highly prone to systematic errors. Our model only contains one spectroscopic redshift, and we are unable to accurately break the mass sheet degeneracy \citep{Schneider:1995vn}. This degeneracy can be broken using multiple source planes, i.e., multiple systems of different source redshift. Although we have included secondary arcs at different source redshifts than the main arc in our model, their model-derived redshifts are inconsistent with photometric redshifts. Our eight lens models produce similar slopes for the mass distribution because they each derive similar redshifts for the secondary arcs. However, because the redshifts may be incorrect (see \S~\ref{subsec:model_redshifts}), the mass sheet degeneracy has been artificially broken; therefore, the mass slopes are likely incorrect. However, the mass enclosed within the critical curve remains the most accurate measurement of the mass.

We estimate the dynamical mass from the radial velocities of cluster member galaxies (\S\ref{sec:specz}) using the $\sigma_{DM}-M_{200}$ scaling relation from \citet{Evrard:2008uo}. The velocity dispersion we measure yields a dynamical mass of $M_{200} = 8.1^{+7.5}_{-5.8}\times10^{14}\ \msol$ for \cluster. With this relation, however, we are assuming that strong lensing clusters are not biased in mass. Strong lensing clusters are more likely to be oriented with the principle axis aligned along the line of sight \citep{Hennawi:2007ek}, resulting in a 19\% bias in cluster mass.

The \citet{Oguri:2012bs} lens model is represented by a single elliptical NFW profile with $M_{vir}=2.26^{+2.41}_{-0.96}\times10^{14}\ \msol$ and concentration parameter $c=22.39^{+17.42}_{-15.70}$. For a circular NFW profile, this yields $M(r<250\ \mathrm{kpc}) = 0.8^{+0.9}_{-0.5}\times10^{14}\ \msol$ and $M(r<500\ \mathrm{kpc}) = 1.3^{+1.3}_{-0.7}\times10^{14}\ \msol$. The two models are in agreement at smaller radii, as both are built from strong lensing. Our model, with more strong lensing constraints and a spectroscopic redshift of the main arc, has a higher-fidelity mass estimate in this region. However, the lack of agreement at larger radii spawns from the lack of weak lensing constraints in our model -- we measure the mass distribution less accurately on the outer regions of the cluster than \citet{Oguri:2012bs}.

\subsection{Magnification}
\label{subsec:magnification}

\begin{figure*}
\includegraphics[scale=0.8,resolution=300]{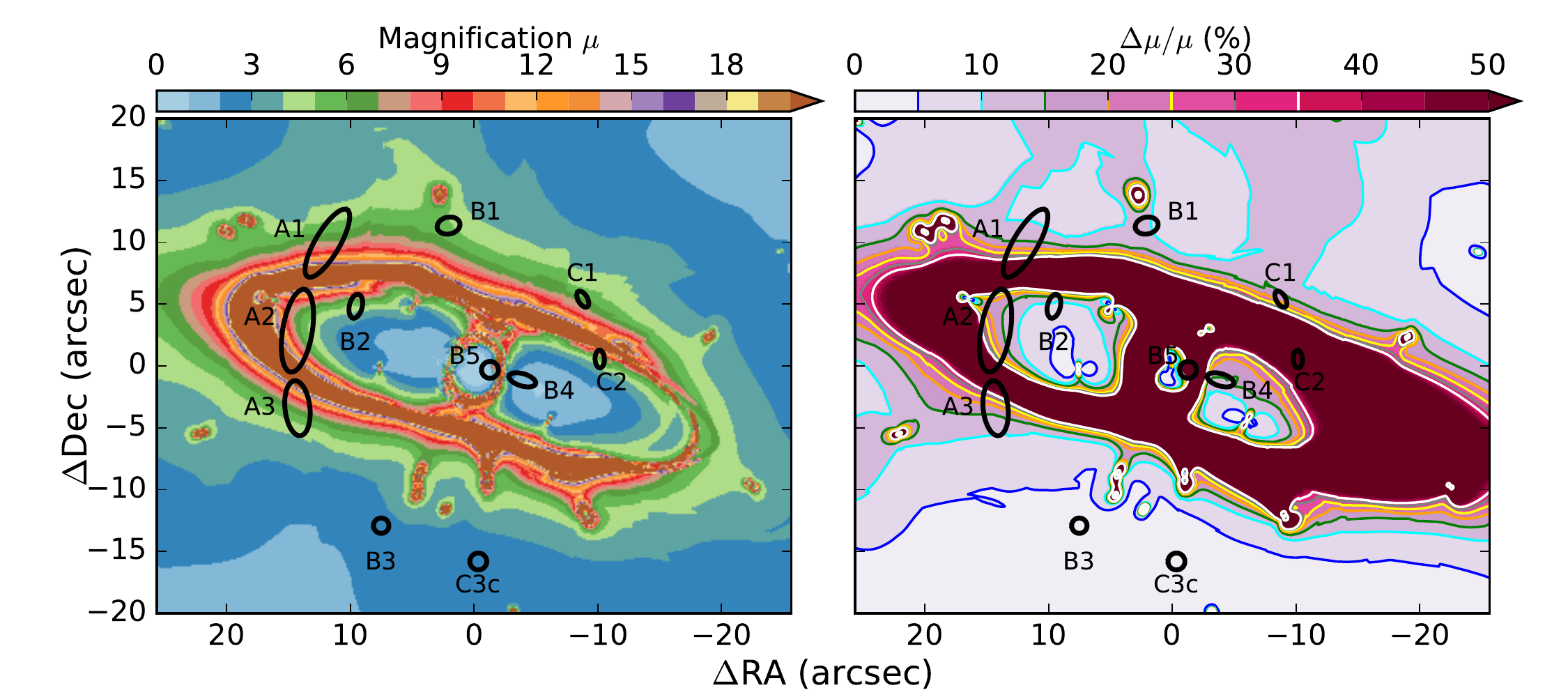}
\caption{The weighted median magnification map for a source at $z=\zA$ stacked from all eight different lens models (left) and the corresponding uncertainty in magnification marginalized across all eight models (right). The locations of the multiple images are shown by the black ovals.}
\label{fig:magmap}
\end{figure*}

We measure the magnification of the arc by combining all magnification maps across the eight lens models. In \autoref{fig:magmap}, we include the median magnification of each pixel in the image plane at $z=\zA$ across all eight lens model realizations, weighted by the errors estimated from the MCMC chain. We find that the magnification across the middle image A2 is $\sim5-10\times$, with a typical statistical error of 20\% for any given pixel. The median and mean pixel count-weighted magnification across the middle image of the arc are $\sim8$ and $\sim9$, respectively.

\subsection{Predicted images for system C}
All of our lens models predict a counter-image for C1 and C2 in the vicinity of our candidate image C3. The barycenter of the image predictions from all eight models is listed in \autoref{tab:arcs}, with an rms scatter of 1\farcs60. This location is only 0\farcs82 away from our candidate image C3, which is well within the scatter of the predictions.

Half of our models (0,1,4,7) predict a set of radial images, C4 and C5, in the center of the cluster opposite of the radial images B4 and B5. The barycenter of the image predictions for the four models that predict these central images are listed in \autoref{tab:arcs}. The predictions for images C4 and C5 are located 4\farcs3 and 0\farcs5 away from the center of the BCG, with a scatter of 0\farcs9 and 0\farcs8, respectively. We are unable to identify any likely candidates for these radial arcs; however, this does not rule out these models. These central images are predicted to be demagnified by a factor of $\sim4-10$, and would be hidden by the light from the BCG; therefore, if they exist, we likely would not see them in this data.

\subsection{Model-predicted redshifts versus photometric redshifts}
\label{subsec:model_redshifts}
Our lens models predict higher redshifts for the image system B than the photometric redshifts of B1/B2/B3 suggest, by more than 5$\sigma$, indicating a high tension between the models and observations. Similarly, the model redshifts of $z\sim3.8$ would lead to a non-detection in F390W, which indicates that the redshift for system C may also be incorrect. Based on observations, the model redshifts are likely incorrect. However, in lens modeling, redshift of the source is not the relevant quantity, but rather the ratio of angular diameter distances between lens and source $d_\mathrm{ls}$ in relation to observer and source $d_\mathrm{s}$. This lensing ratio, $d_\mathrm{ls}/d_\mathrm{s}$, scales the deflection angle, which is what is used to determine the locations of multiple images in the lens modeling process; redshift of a source is a secondary measurement from lens modeling based on choice of cosmological parameters. For system B, a difference in model versus photometric redshifts of $z\sim3.8$ and $z\sim2.7$ equates to only a $\sim10\%$ difference in $d_\mathrm{ls}/d_\mathrm{s}$. This value for the error can be propagated into uncertainties in mass and magnification of sources, as the lensing ratio is used to scale all of the quantities that go into those calculations. Additionally, the \texttt{Lenstool} software has a tendency to bias unknown redshifts used as free parameters to systematically higher values, as investigated in \citet{Johnson:2016rt}, which may help to explain the redshift tension in our models.

\section{Source plane reconstruction for \giantarc}
\label{sec:clump_model}
Gravitational lensing allows us to measure the substructure of galaxies with much higher resolution than field galaxies. Our lens model translates between the observed image plane clump and their physical size and position in the source plane. Naively, the physical size of the clumps can be determined by measuring the image plane area and dividing by the factor of the magnification. This method breaks down quickly when considering unresolved structures, as the true lensed shape of the clump is lost when the lensed image is convolved with the instrument PSF. A more accurate reconstruction of source structure that is at the diffraction limit of the telescope when lensed requires a way to disentangle the effects of nonuniform magnification across the image and instrument PSF.

To this end, we have created a forward modeling technique to reconstruct the sizes of the star-formation clumps in the source plane. The clumps are modeled in the source plane and then ray-traced to the image plane, convolved with the instrument PSF, and compared to the observed data.

Forward modeling techniques, although computationally costly, have been shown in previous lensing studies to be quite useful in accurately reconstructing the source. These techniques have been used frequently with lower-resolution data of sub-millimeter galaxies \citep{,MacKenzie:2014kx,Dessauges-Zavadsky:2016fj}. \citet{Fu:2012yq} reconstruct a parameterized source accounting for very different PSFs/beams from optical to sub-millimeter. With the onset of higher-resolution sub-millimeter facilities, such as the \textit{Atacama Millimeter/Sub-millimeter Array}, these forward-modeling techniques have evolved to allow for a full reconstruction of the source in the complex \textit{uv} plane from interferometric data \citep{Hezaveh:2013vn,Hezaveh:2016ys}.

\subsection{Initial image plane Gaussian decomposition}
Because we are focusing on the clumps, we first perform a Gaussian decomposition of the main arc in order to separate the clumpy structure from the diffuse background. We combine the F606W and F390W to create a higher signal-to-noise detection image. We use GALFIT \citep{Peng:2010qy} to create a parameterized model of the lensed galaxy in the image plane. Two-dimensional Gaussian components are placed in the image plane at the locations of bright clumps and are fit simultaneously to the data.  The best-fit model of the arc is then subtracted from the data to reveal more clumps, which are added to the model and are fit again. This process is done iteratively until the resulting residuals are consistent with the background noise. The final model in F606W+F390W is then used as a template to separately fit each of the F606W and F390W images.

\begin{figure*}
\center
\includegraphics[scale=0.7,resolution=300,trim=0 50 0 50]{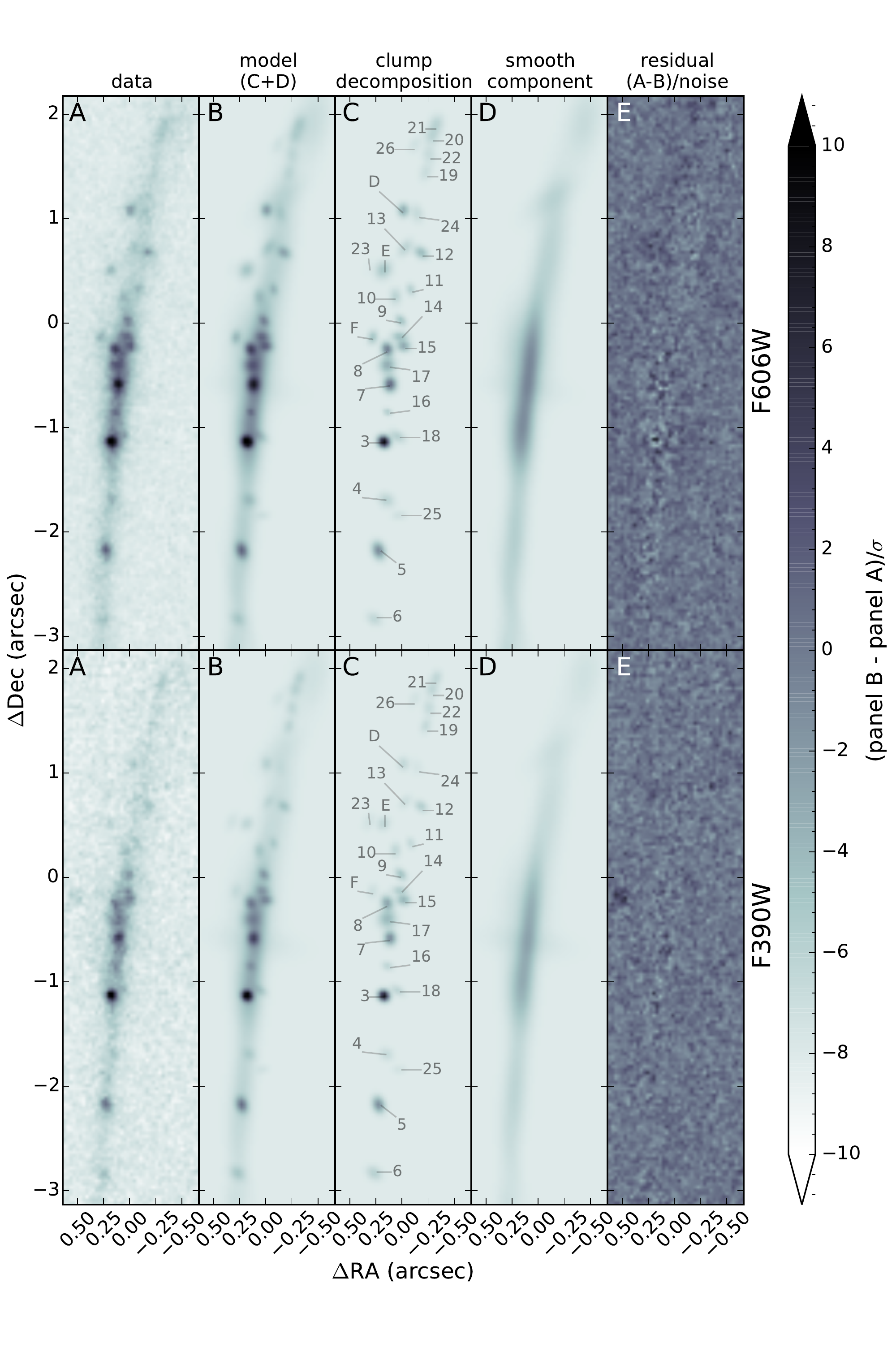}
\caption{GALFIT clump decomposition in F606W (top) and F390W (bottom) imaging. (A) \hst\ imaging of the middle image of \giantarc\ in the image plane. (B) The complete GALFIT model of the middle image of the arc. (C) Clump component of panel B. (D) Smooth component of panel B. (E) Residual of the clump+smooth GALFIT model and the data in units of rms noise.}
\label{fig:galfit_model}
\end{figure*}

We separate the Gaussian components used to create these models into two sets. The clump model consists of bright blobs with sizes roughly a few times that of the \hst\ PSF in the image plane, which will translate to sizes $<100$ pc in the source plane. The smooth model has lower surface brightness and covers nearly the entire length of the arc in the image plane. These models are shown in \autoref{fig:galfit_model}.

\subsection{Source plane clump model}
We model each clump in the source plane as a two-dimensional Gaussian on a grid of 0\farcs003 pixels (to allow for $\sim10\times$ magnification), then ray trace the light distribution back to the image plane via custom ray-tracing code written in Python. We use a Bayesian approach to model the clump parameters, allowing for both parameter space exploration and model comparison. We use the publicly available Python package \texttt{emcee} to perform an affine-invariant sampling of model parameter space \citep{Foreman-Mackey:2013rf}. This sampling provides us with an estimate for the posterior PDF

\begin{equation}
\mathrm{Pr}(\vec\theta | D,M) = \frac{\mathrm{Pr}(D | \vec\theta, M)\ \mathrm{Pr}(\vec\theta | M)}{\mathrm{Pr}(D | M)}
\end{equation}

\noindent for a selection of parameters $\vec\theta$, given the design of the source plane model $M$ and the observed data in the image plane $D$. Here, $\mathrm{Pr}(D | \vec\theta, M)$ is the prior PDF of a $\vec\theta$ for a given model $M$; $\mathrm{Pr}(D | M)$ is the evidence, which normalizes the posterior PDF and accounts for the model complexity; $\mathrm{Pr}(D | \vec\theta, M)$ is the likelihood function of getting the observation $D$, given a source plane model $M$ with parameters $\vec\theta$. According to Bayesian theory, the model that is the best fit will maximize the posterior PDF; one with high likelihood, but is consistent with priors and is more simplistic. Because the evidence is constant for a given M, for this analysis, we will only maximize the non-normalized posterior PDF.

We define the likelihood function of our source plane model as

\begin{equation}
\mathrm{Pr}(D | \vec\theta, M) = \prod_{i=1}^{N} \frac{1}{\sigma\sqrt{2\pi}}\mathrm{Exp}\left[-\frac{\chi^2}{2} \right],
\end{equation}

\noindent where $N$ is the number of pixels in the image plane over the region encompassing the giant arc (see below for definition). The contribution to the overall $\chi^2$ from each image plane pixel is

\begin{equation}
\chi^2  = \sum_{i=0}^{N}\frac{1}{\sigma^2}[I_\mathrm{d}(x_i)-I'_\mathrm{s}(x_i | \vec\theta)]^2,
\end{equation}

\noindent where $I_\mathrm{d}$ is the surface brightness of the observed data at image plane position $x$, and 

\begin{equation}
I'_\mathrm{s}(\vec x | \vec\theta) = I_\mathrm{s}(\vec x | \vec\theta) \ast f(\vec x)
\end{equation}

\noindent is the surface brightness of the model $M$ in the image plane. Here, $I_\mathrm{s}$ is the source plane model surface brightness, ray-traced to the image plane position $x$, which is then convolved in the image plane with the empirical PSF of the instrument $f(x)$.

An empirical PSF is computed for each filter, using data from the entire SGAS \hst\ program GO13003. We select stars in each cluster frame, coadded after subtracting for background features and nearby objects, following \citet{Skelton:2014lr}. We then fit a Gaussian profile to the PSF, and use this as our kernel for smoothing the model to the resolution of \hst.

Because our empirical PSF was averaged over many different epochs of HST observations, it is likely not an exact match to the PSF at the time and position on the detector of the 
SDSS J1110+6459 observations.  According to the WFC3 handbook,\footnote{\url{http://www.stsci.edu/hst/wfc3/documents/handbooks/currentIHB/c06_uvis07.html}} the PSF at 0.4 $\mu$m can vary with breathing by up to 3\%.  Variation of the PSF spatially across the detector can be comparable to the spatial variation.\footnote{\url{http://www.stsci.edu/hst/wfc3/documents/ISRs/WFC3-2013-13.pdf}} Therefore, we include runs of our MCMC that account for a $\pm5\%$ and $\pm10\%$ error in the size of the Gaussian convolution kernel when measuring sizes and fluxes of the clumps. We find no trend in changes of size and flux for the different PSF sizes used, only an overall increase in the statistical errors.

For mapping the image plane to source plane, we use the deflection matrices from Model~2, which has the lowest image plane rms and is close to the median magnification per pixel for the middle image A2 across all eight models. Model~2 produces magnifications across A2 that are neither extremely high nor low compared to the other models. We optimize two parameters for each clump: flux and size. We tested shape parameters (i.e., ellipticity and position angle), but found these parameters could not be constrained for even the brightest and most resolved clumps and thus were not included in the optimization. The clumps are centered on the source plane position that maps to the peak in brightness of that clump in the image plane.  All parameters are assigned uniform random priors.

\begin{figure*}
\includegraphics[scale=0.95,resolution=300]{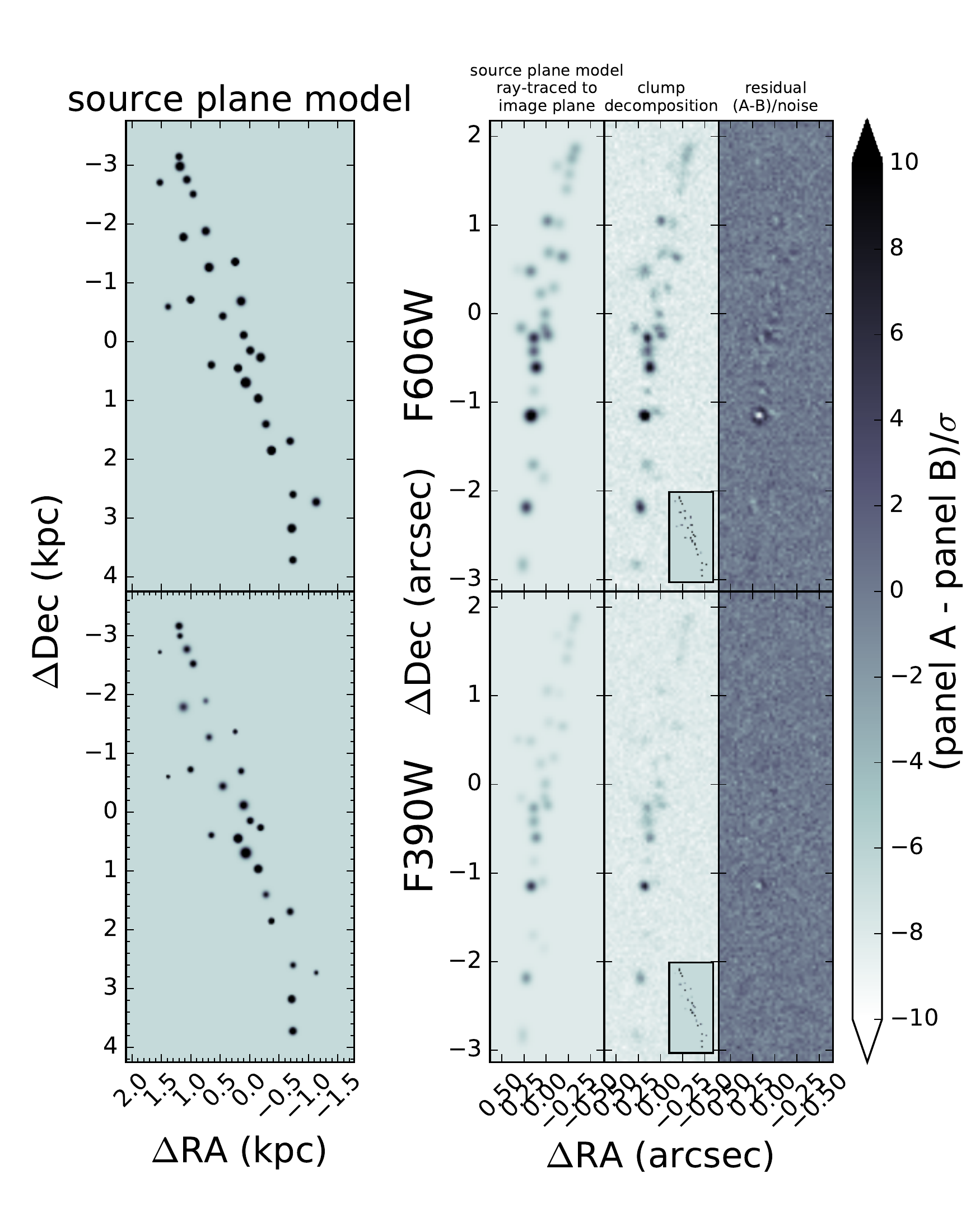}
\caption{Left panel: model of clumps in the source plane. Note: the middle image of the galaxy has negative parity in declination; for display purposes the y-axis has been flipped to match the orientation of galaxy in the image plane. Panel A: the source plane model from the left panel, ray-traced to the image plane and convolved with the \hst\ PSF. Panel B: the clump decomposition model (also Panel B from \autoref{fig:galfit_model} with added noise). The inset shows the source plane model (left panel), scaled to its true angular size with respect to its magnified image. Panel C: residual of source plane model and clump decomposition, in units of rms noise.}
\label{fig:source_plane_model}
\end{figure*}

\begin{figure}
\includegraphics[width=0.5\textwidth,trim={1cm 2.5cm 0.6cm 1cm},clip]{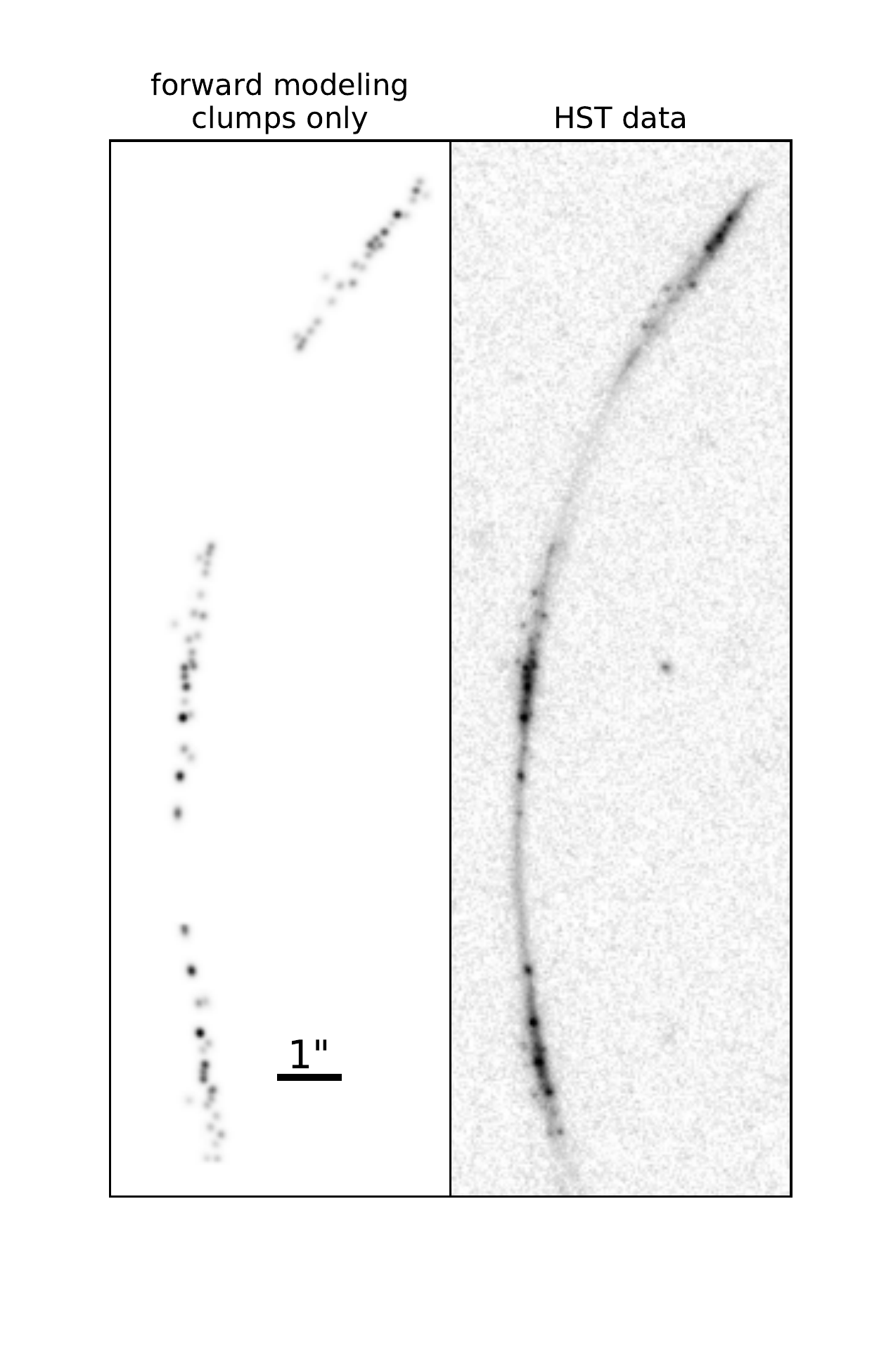}
\caption{Left panel: model of clumps ray-traced to the image plane in all three images of the giant arc. Right panel: \hst\ data of the entire giant arc. Both panels are coadded F606W and F390W. The stretch of the scaling is square root.}
\label{fig:full_arc_model}
\end{figure}

The best-fit clump parameters are given in \autoref{tab:clump_parameters}. We estimate the errors on each parameter from the posterior probability distributions in the MCMC. \autoref{fig:source_plane_model} shows the best-fit model of the source plane clumps in both the image plane and source plane for the central image. \autoref{fig:full_arc_model} shows the clumps ray-traced to all three images of the arc. In this figure, we have removed the clumps corresponding to arcs D, E, and F. The model predicts these images to be at slightly different redshifts than the main arc and will be offset from where they are in A1 and A3. To avoid confusing these clumps with those from \giantarc, we have removed them from the source plane model used in the ray-tracing.

\begin{turnpage}
\capstartfalse
\begin{deluxetable*}{ccccccccccc}
\tablecolumns{7}
\tablecaption{\giantarc\ source plane clump parameters}
\tablehead{\colhead{Clump} & 
                  \colhead{$\Delta$R.A} &
                  \colhead{$\Delta$Decl.} &
                  \colhead{$m_\mathrm{F606W}$} &
                  \colhead{$m_\mathrm{F390W}$} &
                  \colhead{Size} &
                  \colhead{Lensing PSF} &
                  \colhead{Size} &
                  \colhead{Lensing PSF} &
                  \colhead{Magnification} \\[4pt]
	         \colhead{ID} &
	         \colhead{(kpc)} &
	         \colhead{(kpc)} &
	         \colhead{(mag)} &
	         \colhead{(mag)} &
	         \colhead{F606W (pc)} &
	         \colhead{F606W (pc)} &
	         \colhead{F390W (pc)} &
	         \colhead{F390W (pc)} &
	         \colhead{}}
\startdata
D\tablenotemark{$^a$} & 1.08 & -1.71 & 31.83 ($31.76^{+0.10}_{-0.11}$) & 32.81 ($32.54^{+0.15}_{-0.13}$) & 33.6 ($36.6^{+3.2}_{-3.3}$) & $26.8^{+3.1}_{-2.3}$ & 62.3 ($61.5^{+2.6}_{-2.6}$) & $22.1^{+5.4}_{-2.5}$ & 8.5 [5.6-14.2] \\[4pt]
E\tablenotemark{$^a$} & 0.95 & -0.62 & 31.82 ($31.75^{+0.09}_{-0.16}$) & 32.84 ($32.87^{+0.19}_{-0.16}$) & 30.1 ($33.2^{+1.7}_{-1.9}$) & $25.3^{+2.7}_{-1.8}$ & 31.7 ($33.4^{+1.9}_{-2.2}$) & $20.0^{+4.6}_{-2.1}$ & 8.5 [5.5-14.1] \\[4pt]
F\tablenotemark{$^a$} & 0.58 & 0.52 & 32.21 ($32.11^{+0.09}_{-0.14}$) & 33.18 ($32.98^{+0.21}_{-0.18}$) & 30.9 ($32.1^{+2.7}_{-1.7}$) & $27.5^{+3.7}_{-2.8}$ & 36.4 ($37.5^{+2.9}_{-3.0}$) & $18.7^{+6.6}_{-3.4}$ & 9.0 [5.7-14.6] \\[4pt]
3 & -0.50 & 2.01 & 30.44 ($30.45^{+0.04}_{-0.07}$) & 30.75 ($30.73^{+0.08}_{-0.07}$) & 30.5 ($31.8^{+3.5}_{-1.6}$) & $27.9^{+1.8}_{-3.3}$ & 22.0 ($22.1^{+1.8}_{-1.7}$) & $22.6^{+4.3}_{-3.5}$ & 10.2 [6.3-15.6] \\[4pt]
4 & -0.89 & 2.78 & 32.31 ($32.18^{+0.12}_{-0.14}$) & 33.28 ($33.45^{+0.16}_{-0.15}$) & 28.9 ($31.8^{+2.1}_{-1.4}$) & $28.4^{+1.7}_{-2.1}$ & 35.9 ($34.8^{+2.0}_{-1.9}$) & $22.8^{+8.6}_{-4.9}$ & 12.0 [7.3-17.5] \\[4pt]
5 & -0.86 & 3.38 & 31.17 ($31.12^{+0.05}_{-0.11}$) & 31.39 ($31.41^{+0.07}_{-0.07}$) & 32.3 ($33.6^{+1.2}_{-1.2}$) & $27.6^{+2.2}_{-1.6}$ & 32.7 ($34.7^{+2.3}_{-2.1}$) & $24.0^{+8.8}_{-5.2}$ & 15.3 [9.2-21.6] \\[4pt]
6 & -0.88 & 3.93 & 32.32 ($32.30^{+0.15}_{-0.12}$) & 32.31 ($32.65^{+0.16}_{-0.16}$) & 30.7 ($34.8^{+4.2}_{-3.5}$) & $29.1^{+2.6}_{-2.3}$ & 40.3 ($35.8^{+2.5}_{-2.4}$) & $24.3^{+4.2}_{-4.8}$ & 25.0 [14.5-34.7] \\[4pt]
7 & -0.26 & 1.11 & 30.92 ($30.91^{+0.09}_{-0.08}$) & 31.38 ($31.38^{+0.08}_{-0.08}$) & 31.2 ($33.6^{+2.0}_{-2.5}$) & $26.3^{+2.1}_{-3.6}$ & 36.6 ($36.8^{+4.2}_{-3.6}$) & $21.5^{+15.0}_{-4.5}$ & 9.0 [5.6-14.2] \\[4pt]
8 & 0.10 & 0.58 & 31.14 ($31.17^{+0.06}_{-0.07}$) & 31.59 ($31.60^{+0.08}_{-0.08}$) & 29.8 ($32.5^{+2.7}_{-2.1}$) & $27.0^{+2.1}_{-2.9}$ & 43.6 ($38.1^{+4.2}_{-3.8}$) & $20.9^{+5.2}_{-3.6}$ & 8.8 [5.5-14.1] \\[4pt]
9 & 0.00 & 0.00 & 32.28 ($32.23^{+0.09}_{-0.14}$) & 32.35 ($32.22^{+0.12}_{-0.11}$) & 33.1 ($35.0^{+2.6}_{-2.3}$) & $27.7^{+1.8}_{-2.9}$ & 56.9 ($50.3^{+4.2}_{-4.2}$) & $22.9^{+7.4}_{-4.2}$ & 8.2 [5.2-13.3] \\[4pt]
10 & 0.37 & -0.33 & 32.50 ($32.55^{+0.20}_{-0.15}$) & 32.80 ($32.93^{+0.21}_{-0.16}$) & 32.6 ($37.0^{+4.8}_{-3.8}$) & $28.0^{+1.8}_{-1.6}$ & 52.3 ($56.3^{+3.6}_{-3.8}$) & $21.8^{+3.7}_{-3.9}$ & 8.3 [5.3-13.6] \\[4pt]
11 & 0.04 & -0.59 & 32.75 ($32.58^{+0.13}_{-0.13}$) & 33.09 ($32.93^{+0.19}_{-0.16}$) & 48.7 ($42.2^{+4.4}_{-5.3}$) & $28.8^{+1.7}_{-3.5}$ & 38.3 ($40.0^{+3.3}_{-3.3}$) & $22.3^{+6.6}_{-5.3}$ & 7.9 [5.1-13.0] \\[4pt]
12 & 0.15 & -1.28 & 31.90 ($31.89^{+0.08}_{-0.11}$) & 32.83 ($32.84^{+0.18}_{-0.16}$) & 32.9 ($34.4^{+1.7}_{-2.0}$) & $25.7^{+4.3}_{-2.6}$ & 19.9 ($19.6^{+1.9}_{-1.8}$) & $22.3^{+7.6}_{-5.6}$ & 7.9 [5.1-13.0] \\[4pt]
13 & 0.62 & -1.18 & 32.34 ($32.37^{+0.13}_{-0.12}$) & 33.28 ($33.26^{+0.26}_{-0.22}$) & 42.7 ($39.6^{+2.7}_{-3.0}$) & $27.7^{+2.5}_{-3.7}$ & 45.0 ($38.5^{+4.2}_{-4.2}$) & $18.3^{+3.7}_{-4.1}$ & 8.2 [5.3-13.5] \\[4pt]
14 & -0.12 & 0.27 & 32.13 ($32.14^{+0.10}_{-0.11}$) & 32.58 ($32.67^{+0.16}_{-0.14}$) & 33.6 ($34.7^{+2.7}_{-2.9}$) & $25.8^{+2.5}_{-1.7}$ & 36.5 ($38.8^{+2.8}_{-2.9}$) & $20.3^{+9.6}_{-4.4}$ & 8.3 [5.3-13.5] \\[4pt]
15 & -0.30 & 0.39 & 31.72 ($31.72^{+0.15}_{-0.10}$) & 32.14 ($32.08^{+0.10}_{-0.10}$) & 35.9 ($34.2^{+1.7}_{-2.1}$) & $26.9^{+3.4}_{-1.4}$ & 30.4 ($33.8^{+2.7}_{-2.7}$) & $19.8^{+7.4}_{-3.7}$ & 8.3 [5.2-13.3] \\[4pt]
16 & -0.40 & 1.55 & 33.11 ($32.73^{+0.29}_{-0.24}$) & 33.24 ($33.05^{+0.22}_{-0.19}$) & 41.2 ($43.5^{+4.1}_{-4.8}$) & $26.9^{+2.3}_{-2.0}$ & 45.6 ($42.4^{+4.1}_{-4.1}$) & $20.1^{+3.3}_{-2.5}$ & 9.5 [5.9-14.8] \\[4pt]
17 & -0.04 & 0.83 & 31.46 ($31.42^{+0.06}_{-0.06}$) & 31.63 ($31.69^{+0.11}_{-0.09}$) & 42.5 ($42.9^{+4.8}_{-4.4}$) & $26.8^{+2.1}_{-2.2}$ & 63.3 ($54.2^{+5.6}_{-5.0}$) & $23.7^{+6.2}_{-4.2}$ & 8.9 [5.6-14.2] \\[4pt]
18 & -0.83 & 1.85 & 32.92 ($32.88^{+0.24}_{-0.23}$) & 32.75 ($32.97^{+0.22}_{-0.18}$) & 37.2 ($42.2^{+3.6}_{-4.5}$) & $25.7^{+2.2}_{-2.9}$ & 38.7 ($39.0^{+3.1}_{-3.1}$) & $20.4^{+6.8}_{-3.9}$ & 9.6 [5.9-14.6] \\[4pt]
19 & 0.90 & -2.46 & 32.95 ($32.87^{+0.19}_{-0.18}$) & 32.88 ($32.59^{+0.15}_{-0.14}$) & 32.5 ($33.9^{+1.8}_{-2.1}$) & $26.3^{+2.4}_{-2.4}$ & 41.2 ($40.1^{+2.7}_{-3.0}$) & $21.6^{+8.0}_{-5.1}$ & 9.1 [6.0-15.2] \\[4pt]
20 & 1.14 & -2.95 & 32.30 ($32.14^{+0.14}_{-0.17}$) & 33.20 ($33.06^{+0.24}_{-0.21}$) & 44.1 ($37.2^{+4.9}_{-6.2}$) & $27.7^{+1.9}_{-2.6}$ & 30.9 ($30.2^{+3.9}_{-3.7}$) & $19.7^{+5.1}_{-3.1}$ & 10.3 [6.9-17.3] \\[4pt]
21 & 1.16 & -3.12 & 32.39 ($32.21^{+0.12}_{-0.16}$) & 32.54 ($32.70^{+0.17}_{-0.15}$) & 30.7 ($32.9^{+3.2}_{-2.3}$) & $26.8^{+1.9}_{-2.2}$ & 38.2 ($31.1^{+3.1}_{-2.9}$) & $20.2^{+5.3}_{-3.9}$ & 10.8 [7.3-18.3] \\[4pt]
22 & 1.02 & -2.71 & 32.88 ($32.60^{+0.22}_{-0.22}$) & 32.90 ($32.92^{+0.21}_{-0.17}$) & 39.0 ($47.6^{+4.4}_{-7.3}$) & $26.9^{+2.9}_{-3.3}$ & 51.3 ($42.2^{+4.5}_{-4.4}$) & $24.0^{+3.5}_{-5.4}$ & 9.5 [6.4-16.0] \\[4pt]
23 & 1.35 & -0.50 & 33.83 ($33.44^{+0.54}_{-0.31}$) & 33.49 ($33.43^{+0.26}_{-0.23}$) & 37.4 ($35.4^{+2.0}_{-2.1}$) & $28.7^{+1.8}_{-1.9}$ & 16.8 ($16.4^{+2.3}_{-2.2}$) & $23.1^{+4.7}_{-7.7}$ & 8.9 [5.8-14.7] \\[4pt]
24 & 0.68 & -1.82 & 32.93 ($32.73^{+0.26}_{-0.18}$) & 33.99 ($33.75^{+0.42}_{-0.31}$) & 45.2 ($42.0^{+3.6}_{-2.8}$) & $27.4^{+2.5}_{-1.5}$ & 40.9 ($40.8^{+4.7}_{-5.1}$) & $24.3^{+5.5}_{-6.8}$ & 8.3 [5.5-13.8] \\[4pt]
25 & -1.30 & 2.91 & 33.13 ($33.02^{+0.25}_{-0.24}$) & 33.96 ($33.38^{+0.35}_{-0.27}$) & 49.3 ($46.3^{+4.5}_{-4.0}$) & $27.8^{+1.4}_{-2.0}$ & 27.5 ($35.4^{+4.9}_{-4.8}$) & $24.1^{+4.1}_{-3.4}$ & 11.8 [7.2-17.2] \\[4pt]
26 & 1.50 & -2.67 & 33.22 ($32.80^{+0.21}_{-0.22}$) & 33.88 ($33.51^{+0.32}_{-0.29}$) & 32.5 ($33.4^{+2.6}_{-2.5}$) & $25.2^{+2.1}_{-1.3}$ & 18.2 ($21.9^{+4.9}_{-4.4}$) & $25.4^{+5.9}_{-8.3}$ & 10.0 [6.7-16.8]
\enddata
\tablenotetext{$^a$}{These ``clumps" correspond to sources D, E, and F in the lens model, and are not part of the galaxy. We report their sizes and fluxes at the redshift of the main arc, $z=\zA$.}
\tablecomments{Column (1) clump Identifier; (2) and (3) relative R.A and decl., in kpc at arc redshift, relative to clump \#9; (4) F606W magnitude of clumps in the source plane; (5) same as (4), for F390W; (6) source plane size (HWHM) of the clumps in F606W, in parsecs; (7) size of lensing PSF in the source plane in F606W, in parsecs; (8) and (9) same as (6) and (7), respectively, for F390W; (10) clump magnification. For columns 4, 5, 6, 8, we report the best-fit parameter first, and in parenthesis show the median and $1\sigma$ errors on the parameter from the MCMC. For columns 7 and 9, we report the  median and $1\sigma$ errors on the parameter from the MCMC. The last column shows the median magnification; in brackets, the full range of magnifications is given for all eight lens models.}
\label{tab:clump_parameters}
\end{deluxetable*}
\capstarttrue
\end{turnpage}

\subsection{Completeness Analysis}
We expect our results to be affected by observational biases, in that fainter and smaller clumps are less likely to be detected. Given that the sizes of star-forming regions in the local universe follow a power law, we expect there to be many more of these difficult to detect clumps in galaxies \citep{Kennicutt:1989fk}; thus, our models are incomplete.

To understand our completeness limits, we run simulations determining the efficiency of detection of clumps, based on their size and flux. We create a set of 1000 lensed galaxies similar in design to \giantarc, using new parameters for the simulated clumps. At each image plane position of the clumps in our model of \giantarc, we assign a new position perturbed by a few pixels. We ray-trace all new positions for the simulated clumps to the source plane, where we place fake clumps. The parameters for 18 (2/3) of those fake clumps are drawn randomly from the list of best-fit parameters of the detected clumps, with replacement (i.e., the parameters listed in \autoref{tab:clump_parameters}). For the remaining 9 (1/3) clumps, we select parameters randomly to have a $m_\mathrm{F606W}=30-37$ and $r=1-40$ pc. These ranges were chosen to cover the parameter space where we expect to measure a significant change in the efficiency function for detecting a clump. For simplicity, we assign all clumps the same color $m_\mathrm{F390W}-m_\mathrm{F606W}=0.36$, which is the typical color of the clumps derived from the source plane measurements from the forward modeling MCMC. All the clumps have the same size in both F606W and F390W. We then ray-trace the source plane models for both F390W and F606W of the fake clumps back to the image plane, convolve with their respective PSFs, and coadd the images. We then add the model of the F606W+F390W smooth component to the fake clumps and add realistic noise.

Next, we run our clump-finding algorithm on the simulated lensed galaxies, where we create a GALFIT model of the entire galaxy, clumps and smooth component. The only inputs used for the algorithm are the new image plane positions of the fake clumps. Added to the GALFIT model are three locations that are not included in the simulated galaxy, which we select randomly from three of the exact positions of the clumps chosen. The purpose of these three additional components is to determine the typical background level of the clumps at those positions. GALFIT will attempt to fit a clump at those locations, even though there is no assigned flux in the source plane model that maps to that position. The output magnitude of that false clump tells us the position-dependent threshold for whether or not a clump near that position can be detected. This threshold is influenced by many factors: sky background, magnification, the brightness of the smooth component at that position, and nearby clumps that may overlap. We define the background level at each position to be the median magnitude of the false clumps measured by GALFIT. We define a simulated clump as ``detected" if its GALFIT magnitude is brighter than one standard deviation above the background level at the position where it was measured. All the clumps were detected well above this limit.

Our simulations reveal that clump detection depends strongly on the flux of the clump in F606W, and is independent of size for clumps that are larger than 10 pc in the image plane. As we will show below, our resolution limit is roughly 20 pc. Therefore, we combine the efficiency measurements across all clumps larger than 10 pc, and fit the efficiency as a function of magnitude using

\begin{equation}
\eta(m) = \frac{N_0}{1 + \exp[(m-m_{0.5})/s]}.
\label{eqn:efficiency}
\end{equation}

\begin{figure}
\label{fig:efficiency_fits}
\includegraphics[scale=0.4,resolution=300]{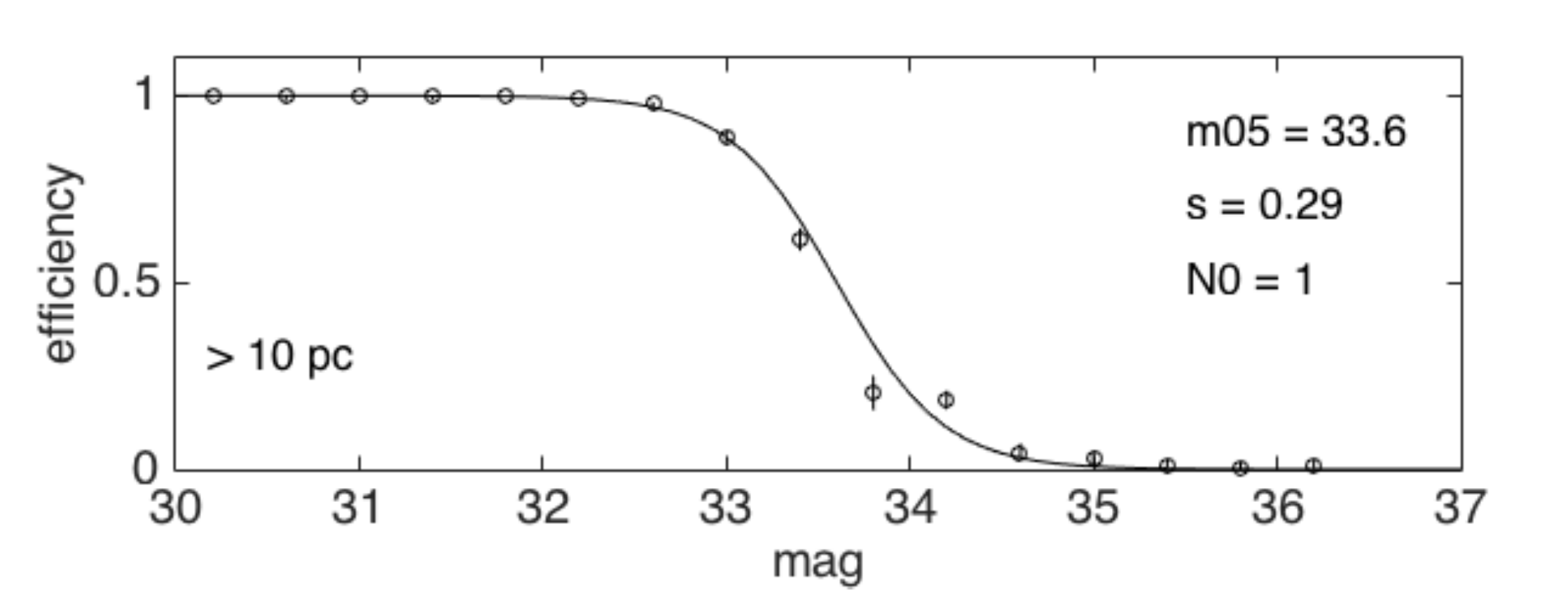}
\caption{Model fits to the empirical efficiency for clumps larger than 10 pc, as a function of clump magnitude. The best-fit parameters to \autoref{eqn:efficiency} for each size bin are shown in the upper right-hand corner.}
\end{figure}

\noindent The model fit and parameters are shown in \autoref{fig:efficiency_fits}. Our 80\% completeness limit is 33.2 mag.

Although our detection efficiency depends only on clump magnitude, our model is still limited in the size of clumps measured, due to resolution limits. To determine the size limit, we create a model of \giantarc\ in the image plane, where we have replaced each clump with the instrument PSF, the highest resolution we can achieve for a given clump at that location within the galaxy. Each of these PSFs are given a uniform brightness that roughly matches the average measured brightness of all clumps in the image plane. We apply our forward modeling algorithm to this model, to determine the size of the lensing PSF in the source plane; i.e., the smallest size we can measure in the source plane, given its magnification and the instrument PSF. The sizes of the lensing PSFs for the clumps in F390W and F606W are given in \autoref{tab:clump_parameters}, ranging from 24 to 31 parsecs for F606W and from 17 to 30 parsecs in F390W.

\section{Summary}
\label{sec:summary}

\subsection{Hybridization of lens modeling}
In this paper, we have demonstrated that a multiscale grid model for the mass distribution, which was first implemented for Abell 1689 by \citet{Jullo:2009ij}, can be applied to smaller lensing clusters like \cluster. We attempted to model \cluster\ with a traditional parametric lens model, and found that it was impossible to robustly reconstruct the source plane surface brightness of \giantarc\ robustly in all three images. Adding the additional flexibility of a multiscale grid allows for a model that accurately reconstructs the source galaxy. Quantitatively, the image plane rms decreased from 1\farcs4 in the parametric model to typically 0\farcs1 in the multiscale grid models.

\subsection{Advantages and disadvantages of forward modeling versus traditional ray-traced source plane reconstruction}
\label{subsec:method_comparison}

Our forward modeling methodology has allowed us to obtain unprecedented physical resolution of galactic structure of a galaxy at $z=\zA$ in the source plane. Field galaxies (i.e., unlensed) were typically resolved down to one kiloparsec scales in surveys with \hst. Previous studies of lensed galaxies have uncovered resolution limits on the order of a few hundred parsecs \citep{Jones:2010uq,Livermore:2012fk,Swinbank:2012jk,Wisnioski:2012qf,Livermore:2015ve}. Our methodology effectively allows us to deconvolve the source plane structure with the lensing PSF, which is the effect of applying the instrument PSF in the image plane to a galaxy that is magnified asymmetrically. The magnification $\mu$ of an object is defined as the ratio of image plane area to source plane area; therefore, if the lensing shear is isotropic, the ratio of the image plane radius and source plane radii of a circular object should be equal to the square root of the magnification, $\sqrt\mu$. In cases where the shear is not isotropic, as is typically the case in giant arcs like \giantarc\ (where the galaxy is lensed tangentially around the center of the lensing potential), we expect the ratio between the radius in the image plane and the semi-minor (semi-major) axis in the source plane to be slightly larger (smaller) than $\sqrt\mu$. In the image plane, the smallest resolvable angular scale is determined by the instrument PSF. The \hst\ F606W PSF has a Gaussian width of 0\farcs033 (FWHM=0\farcs078); thus, the smallest resolvable physical scale in the source plane would correspond to roughly $0\farcs033/\sqrt\mu$. For a source a $z=\zA$ and $\mu=12$ (median magnification of the clumps), this scale corresponds to roughly 77 pc. Therefore, we expect that ray tracing \giantarc\ in the usual manner would not be able to measure sizes smaller than about 60-70 pc. However, our results reveal that \giantarc\ does not have clumps larger than 40 pc. 

We directly compare our forward modeling results to the traditional ray-tracing methodology. We create a source plane model of the clumps by ray-tracing the GALFIT image plane model of the clumps in F606W back to the source plane, using methods similar to \citet{Sharon:2012ly}. We then measure the size of each clump in the source plane by fitting a two-dimensional Gaussian centered on the position of each clump in the model, with four free parameters: amplitude, semi-major axis, semi-minor axis, and position angle. Our Gaussian fits to our ray traced clumps in \giantarc\ have a median semi-minor (semi-major) axis fit of 86.9 (143) pc, which is over twice the value of the largest clump we measure using our forward modeling technique. We find that, for individual clumps, the highest resolution achieved by ray tracing (i.e., the semi-minor axis) is $3.5\pm1.6$ times larger than that measured through forward modeling. Our methods allow us to obtain higher resolution isotropically, rather than along the axis of the clump that is tangential to the direction of the shear. We measure a typical axis ratio (semi-major axis/semi-minor axis) to be $1.8\pm0.8$, usually with this tending toward higher values for higher-magnification clumps. Therefore, the semi-major axis of the clumps will still be measured as $5.6\pm1.8$ times larger than the forward modeling measurement.

The results show that forward modeling produces a much higher resolution view of source plane structure compared to traditional ray tracing, especially when measuring structure smaller than the size of the ray-traced PSF in the source plane. However, what traditional ray tracing lacks in spatial resolution, it gains over forward modeling in computational speed. Ray-tracing pixels from an image plane grid to a source plane grid takes a matter of minutes and needs only be done once for a single image of a source; the same solution to the lensing equation can be applied to any image plane surface brightness model created on the same grid of pixels. The forward modeling method takes advantage of the same procedure as ray-tracing does, in that it maintains the same solution for mapping source plane to image plane pixels; however, it needs to be run many times for full parameter space exploration. This technique simplified the surface brightness profile of the source galaxy with two-parameter Gaussian profiles representing each of 27 clumps. A single model can be produced in under 2 s, but the 300,000 models produced in the MCMC take roughly 16 hr to complete (on four cores with a 2.20 GHz processor).

The speed of forward modeling is thus limited in how many free parameters are included in the source plane model. To model the source galaxy in more detail would require more parameters, or ideally, a non-parametric approach where each the brightness of each source plane pixel is its own free parameter in the model. This non-parametric approach can be developed for future work regarding high-resolution studies of lensed galaxies, either by increasing the computational resources beyond those we used in this work, or repurposing adaptive mesh refinement codes to work for lensing.

\subsection{Magnification uncertainties}
The systematic uncertainty on magnification is the most significant uncertainty in measuring the sizes of the lensed clumps in the source plane. As we found in \S~\ref{subsec:magnification}, the eight models that we produced for this cluster produce median magnifications across the giant arc \giantarc\ ranging from 18 to 36. A higher or lower typical magnification of a model will shift the size distribution of the clumps by roughly a factor of $1/\sqrt{\mu}$. Therefore, a $\sim60\%$ systematic error on magnification translates to a systematic error of $\sim20\%$ on the source plane sizes of the clumps.

\section{Conclusion}
\label{sec:conclusion}

We have used the power of \hst\ imaging and strong gravitational lensing to resolve structure on $<100$ pc scales in a lensed galaxy \giantarc\ at $z=\zA$. The mass distribution of the lensing galaxy cluster \cluster\ at $z=\zlens$ was mapped through a hybrid parametric-non parametric lens modeling technique developed specifically for this investigation. We measured spectroscopic redshift for the main arc and cluster member galaxies, which fixes the lensing geometry of the lens equation and provides a more robust estimate for the surface mass density of the cluster. We find that our strong lensing mass estimate is consistent with a dynamical mass estimate measured from the velocity dispersion of the cluster, as well as previous strong+weak lensing models of this cluster performed without spectroscopic data. From the lensing mass, we determine the deflection tensors that provide the translation between the observations of the lensed galaxy made in the image plane and the true surface brightness distribution of the galaxy in the source plane. We model the central, most highly magnified image of the lensed galaxy with GALFIT, decomposing the clumpy component from the smooth distribution of light, and implement a forward modeling technique to measure the sizes and luminosities of the clumps in the source plane. Our completeness analysis shows that we have detected the vast majority of clumps brighter than 33.2 mag in F606W, and have achieved a typical resolution limit of $\sim$20-30 pc (magnification dependent) across the galaxy.

Our study has demonstrated the usefulness of gravitational telescopes for understanding the structure of galaxies during the peak epoch of universal star formation. Exciting as these studies are, current sample sizes of lensed galaxies are too small to generalize to the entire galaxy population at high redshift. Many giant arcs have been discovered through various surveys; the bottleneck of the analysis remains in developing accurate lens models to robustly reconstruct the galaxies in the source plane. Strong lens modeling is far from an automated process; identifying multiple images, measuring spectroscopic errors, and computing the models requires considerable human effort for each lensing cluster. Additionally, these studies require a full analysis of the systematic errors of the modeling process. Strong lensing systematics are currently being studied in the context of the most massive, most effective lenses, i.e., the Frontier Fields \citep[see][]{Meneghetti:2016xe,Johnson:2016rt}; however, the parameter space relevant to this work remains unexplored: low-mass clusters with few multiple images and even fewer spectroscopic redshifts. Although the most well-studied lensing clusters are among the most massive in the universe, the majority of clusters that produce giant arcs are those that are most common: low-mass clusters. These clusters have smaller lensing cross-sections, and therefore will typically lens fewer multiple image systems that can be used as constraints. Thus, it is imperative that lensing systematics be studied in small cluster systems, so that future studies similar to this work on \giantarc\ to have the highest accuracy.

We will enter deeper discussions on the scientific impact of the clump sizes and brightnesses we have measured in this paper in future work. Paper II will show how high magnification is necessary to reveal structure within galaxies at high redshift, as we will show in a comparison of our resolved model of \giantarc\ compared to our model of this galaxy, mocked to the resolution and depth of the CANDELS survey. In Paper III, we will analyze the size and brightness distributions of the clumps, and compare our results for the surface density of star formation with those of other galaxies across cosmic time.

\acknowledgements
Support for program \# 13003 was provided by NASA through a grant from the Space Telescope Science Institute, which is operated by the Association of Universities for Research in Astronomy, Inc., under NASA contract NAS 5-26555. We thank the Spitzer Science Center for prompt, detailed, expert advice on reducing our data. K.E.W. gratefully acknowledges support by NASA through Hubble Fellowship grant \#HF2-51368 awarded by the Space Telescope Science Institute, which is operated for NASA by the Association of Universities for Research in Astronomy, Inc., under contract NAS 5-26555.


\end{document}